\documentclass{aa}  

    \usepackage{graphicx}
    \usepackage{txfonts,textcomp}
    \usepackage{natbib,multirow} 
    \usepackage{hyperref}
    \usepackage{orcidlink}
    
\begin{document}

       \title{The SPIRou Legacy Survey}
    
       \subtitle{Detection of a nearby world orbiting in the habitable zone of Gl\,725\,B achieved by correcting strong telluric contamination in near-infrared radial velocities with \texttt{wapiti}}
    \titlerunning{Detection of a nearby world orbiting in the habitable zone of Gl\,725\,B with SPIRou}
    
    \author {M. Ould-Elhkim \inst{1}
              \fnmsep\orcidlink{0000-0002-1059-0193}
              \and
              C. Moutou \inst{1}\orcidlink{0000-0002-2842-3924}
              \and
              J-F. Donati \inst{1}\orcidlink{0000-0001-5541-2887}
              \and 
               P. Cort\'es-Zuleta \inst{2}\orcidlink{0000-0002-6174-4666}
              \and
              X. Delfosse \inst{3}
              \and
              \'E. Artigau \inst{4,5}
              \and
              C. Cadieux \inst{4}
              \and
              P. Charpentier \inst{1}
              \and
              A. Carmona \inst{3}
              \and
              I. Boisse \inst{6}
              \and
              C. Reylé \inst{7}
                \and
              E. Gaidos \inst{8}
                \and
                R. Cloutier\inst{9, 10}\orcidlink{0000-0001-5383-9393}
                  \and
              G. Hébrard \inst{11}
              \and
              L. Arnold \inst{12}
              \and
           J.-D. do Nascimento, Jr. \inst{13}
            \and
              N.J. Cook \inst{4}
            \and 
              R. Doyon \inst{4}
              }
    
       \institute{Universit\'e de Toulouse, UPS-OMP, IRAP, CNRS, 14 avenue E. Belin, Toulouse, F-31400, France
        \and  
        SUPA School of Physics and Astronomy, University of St Andrews,
        North Haugh, St Andrews KY16 9SS, UK
        \and
        Univ. Grenoble Alpes, CNRS, IPAG, 38000 Grenoble, France
        \and  
        Trottier Institute for Research on Exoplanets, Université de Montréal, Département de Physique, C.P. 6128 Succ. Centre-ville, Montréal,
        QC H3C 3J7, Canada
        \and 
        Observatoire du Mont-M\'egantic, Universit\'e de Montr\'eal, D\'epartement de Physique, C.P. 6128 Succ. Centre-ville, Montréal, QC H3C 3J7, Canada
        \and
        Aix Marseille Univ, CNRS, CNES, Institut Origines, LAM, Marseille, France
        \and
        Universit\'e Marie et Louis Pasteur, CNRS, Observatoire des Sciences de l'Univers THETA Franche-Comt\'e Bourgogne, Institut
    UTINAM (UMR 6213), F-25000 Besançon, France
    \and
    Department of Earth Sciences, University of Hawaii at Manoa, Honolulu, Hawaii 96822 USA
        \and 
        Center for Astrophysics | Harvard \& Smithsonian, 60 Garden Street, Cambridge, MA 02138, USA
        \and 
        Department of Physics \& Astronomy, McMaster University, 1280 Main Street West, Hamilton, ON L8S 4L8, Canada
        \and
        Institut d'astrophysique de Paris, UMR7095 CNRS, Universit\'e Pierre \& Marie Curie, 98bis boulevard Arago, 75014 Paris, France 
        \and
        Canada France Hawaii Telescope (CFHT) Corporation, UAR2208 CNRS-INSU, 65-1238 Mamalahoa Hwy, Kamuela 96743 HI, USA
        \and
    Universidade Federal do Rio Grande do Norte (UFRN), Departamento de Física, 59078-970, Natal, RN, Brazil
        }

    % \abstract{}{}{}{}{} 
    % 5 {} token are mandatory
     
      \abstract
        {M dwarfs are prime targets in the search for exoplanets because of their prevalence and because low-mass planets can be better detected with radial velocity (RV) methods. In particular, the near-infrared (NIR) spectral domain offers an increased RV sensitivity and potentially reduced stellar activity signals. Precise NIR RV measurements are strongly affected by telluric absorption lines from the Earth's atmosphere, however.}
        {We searched for planets orbiting Gl\,725\,B, a nearby late-M dwarf located at 3.5\,pc, using high-precision SPIRou RV observations. We also assessed the effect of telluric contamination on these measurements and evaluated the performance of the weighted principal component analysis reconstruction (\texttt{wapiti}) method, which is a weighted principal component analysis (wPCA) approach designed to mitigate these systematics and to improve the sensitivity of planet detections.}
       {Using synthetic and observational SPIRou data, we simulated the effect of telluric lines on RV data under varying barycentric Earth radial velocity (BERV) conditions. We then applied the \texttt{wapiti} method for identifying and correcting telluric-induced systematics in line-by-line RVs. The method was tested through an injection-recovery test on simulated data and was subsequently applied to real SPIRou observations of Gl\,725\,B.}
        {\texttt{wapiti} successfully corrects telluric contamination in simulated and real datasets. This enhances the detectability and accuracy of planetary signals. In the corrected Gl\,725\,B dataset, we identified a two-planet system composed of a candidate inner planet (Gl 725 Bb), with periods of $4.765 \pm 0.004$ days and a semi-amplitude of $1.4 \pm 0.3$\,m$\cdot$s$^{-1}$, and {a confirmed planet} Gl\,725\,Bc, with a period of $37.90 \pm 0.17$ days and a semi-amplitude of $1.7 \pm 0.3$\,m$\cdot$s$^{-1}$. Their minimum mass is $1.5 \pm 0.4$\,M$_\oplus$ and of $3.5 \pm 0.7$\,M$_\oplus$, respectively, and the outer planet is located in the habitable zone of its host star. Using a multi-dimensional Gaussian process framework to model and correct for stellar activity, we also recovered a stellar rotation period of $105.1 \pm 3.3$ days.}      
           {}
    
       \keywords{Techniques: radial velocities;  Techniques: spectroscopic;  Methods: data analysis; Stars: individual: Gl\,725\,B; Stars: planetary systems; Stars: low-mass}
    
       \maketitle
    
    \section{Introduction}
    
    High-precision radial velocity (RV) is a powerful technique for detecting and characterizing exoplanets by measuring their masses and orbital parameters. The accuracy of RV measurements is affected by photon noise, instrumental stability, telluric contamination, and stellar activity \citep{plavchan, fischer}. Stellar surface inhomogeneities such as spots and plages can introduce RV signals of a few m$\cdot$s$^{-1}$ \citep{dumusqueI, dumusquelbl, cegla}. These activity-induced variations depend on the wavelength and are often reduced in the red, optical, and NIR \citep{Marchwinski, robertson, Cale}, at least for the most active M dwarfs \citep{pia}. This makes multiwavelength RV studies important for distinguishing between stellar and planetary signals \citep{huelamo, mahmud, carmona2023nearir, pia}.
    
    \par 
    
    The interest in M dwarfs has grown with the realization that they host more rocky planets than solar-type stars \citep{Bonfils_2013, DressingCharbonneau2015, 2016MNRAS.457.2877G, 2020MNRAS.498.2249H, 2021A&A...653A.114S, Mignon2025}. Their lower masses enhance RV signals from small planets \citep{proximacenb, proxima, gj1002}, and they are both numerous and close to us \citep{henry}. Moreover, the flux contrast between active regions and the photosphere is typically lower than in Sun-like stars, which might reduce the RV jitter induced by stellar activity \citep{Berdyugina, Crockett, carmona2023nearir, pia}, especially when this is combined with variations in line depths between spots and photospheres \citep{Larue_2025}.
    
    \par 
    
    These favorable characteristics have motivated the development of a new generation of NIR spectrographs, such as GIANO \citep{2016ExA....41..351C}, HPF \citep{Metcalf2019}, SPIRou \citep{donati2020}, CARMENES-NIR \citep{bauer2020}, IRD \citep{hirano2020}, iSHELL \citep{Rayner2012}, NIRPS \citep{2022SPIE12184E..54T}, and CRIRES+ \citep{dorn2023}. It remains difficult to achieve high RV precision in the NIR, however, because of telluric absorption and emission lines from the Earth’s atmosphere \citep{bean2010}. These lines are mainly due to water vapor, methane, carbon dioxide, and oxygen, and they are more prominent in the NIR and do not share the stellar Doppler shift.
    
    \par 
    
    Telluric lines have traditionally been corrected for using observations of hot featureless stars \citep{artigau2014}, and water line residuals as low as 2\% were achieved in the 900–1350\,nm range \citep{Sameshima}. More recent approaches relied on synthetic telluric models such as \texttt{TAPAS} \citep{tapas}, with best-case residuals at 1–2\% \citep{ulmer}.
    
    \par 
    
    The residual telluric contamination can still significantly affect the RV precision, especially for targets with poor coverage of the barycentric Earth radial velocity (BERV; i.e., a BERV span $\leq 10$\,km$\cdot$s$^{-1}$ over a year). The BERV corresponds to the projection of the Earth's velocity vector onto the line of sight to the star, and its amplitude varies depending on the stellar declination and meridian transit. Poor BERV coverage limits the modulation of telluric lines in observations and makes it harder to distinguish them from true stellar signals. Simulations performed by \citet{tell1, tell2} using synthetic spectra have shown that with optimized mitigation, the RV scatter from telluric residuals can be reduced to below 1.5\,m$\cdot$s$^{-1}$ in the NIR.
    
    \par 
    
    In this context, the case of Gl\,725\,B is particularly relevant. This nearby late-type M dwarf is a target of the SPIRou Legacy Survey Planet Search (SLS-PS\footnote{\url{https://spirou-legacy.irap.omp.eu/doku.php/}}) program, which aims to detect and characterize planetary systems around $\sim$50 M dwarfs using high-precision NIR RV measurements \citep{2023arXiv230711569M}. The source Gl\,725\,B has a narrow BERV coverage of less than 5\,km$\cdot$s$^{-1}$, however, which limits our ability to separate the effects of telluric contamination over time. 
    
    \par 
    
    We simulate the effect of telluric absorption on SPIRou RV data specifically for Gl\,725\,B using the SPIRou Data Reduction Software \texttt{APERO} \citep{apero} and a line-by-line (LBL) RV extraction method \citep{dumusquelbl, etiennelbl}. We then assess the capability of the \texttt{wapiti} (Weighted principAl comPonent analysIs reconsTructIon; \citealt{wapiti}) method to mitigate telluric systematics in these simulations. Finally, we apply \texttt{wapiti} to real SPIRou observations of Gl\,725\,B, which leads to the detection of a two-planet system, including a candidate that is located within the habitable zone.
    
    \par 
    
    The structure of this paper is as follows. Section \ref{sec:gl725b} introduces the M dwarf Gl\,725\,B and presents its SPIRou RV and TESS photometric data. Section \ref{sec:simulation} describes our simulation setup and the effect of limited BERV coverage. In Sect. \ref{sec:wapiti} we perform an injection-recovery test to evaluate the performance of \texttt{wapiti}. Section \ref{sec:725b} presents the analysis of SPIRou observations of Gl\,725\,B and reports the detection of a two-planet system. We summarize our conclusions in Sect. \ref{sec:ccl}.
    
    %--------------------------------------------------------------------
    \section{Observations of Gl\,725\,B}\label{sec:gl725b}
    
    The M5.0V dwarf star Gl\,725\,B (Struve 2398 B, HD 173740)  \citep{kirkpatrick} lies at a distance of $3.5231 \pm 0.0003$ pc from Earth, {as reported by \citet{gaia}}. A summary of its physical properties is available in Table \ref{gl725b_stellar_properties}. Additionally, the longitudinal magnetic field ($B_\ell$) of this star using circularly polarized spectropolarimetry was analyzed by \citet{donatibell} and led to the identification and characterization of a rotational period for Gl\,725\,B, estimated to be $135 \pm 15$ days.
    
\begin{table}
    \centering
    \caption{Stellar parameters of Gl\,725\,B}
    \begin{tabular}{l c c}
    \hline
    \noalign{\smallskip}
    Parameter & Gl\,725\,B & Ref. \\
    \noalign{\smallskip}
    \hline
    \noalign{\smallskip}
    $T_{\mathrm{eff}}$ (K) & $3379 \pm 31$ & 1\\
    {[M/H]} & $-0.28 \pm 0.10$ & 1\\
    {[$\alpha$/Fe]} & $0.14 \pm 0.04$ & 1\\
    $\log g$ & $4.82 \pm 0.06$ & 1\\
    Radius ($R_\odot$) & $0.280 \pm 0.005$ & 1\\
    Mass ($M_\odot$) & $0.25 \pm 0.02$ & 1\\
    $\log (L/L_\odot)$ & $-2.038 \pm 0.003$ & 1\\
    $P_{\mathrm{rot}}$ (d) & $135 \pm 15$ & 2\\
    $P_{\mathrm{rot}}$ (d) & $105.1 \pm 3.3$ & This work\\
    Age (Gyr) & $8.7 \pm 2.1$ & This work\\
    \noalign{\smallskip}
    \hline
    \end{tabular}
    \vspace{0.5em}

    {\footnotesize \textbf{References:} 
    \textbf{1}: \citet{cristofari}; 
    \textbf{2}: \citet{donatibell}.}
    \label{gl725b_stellar_properties}
\end{table}

    The source Gl\,725\,B is part of a binary system. Its companion Gl\,725\,A is classified as an M3V dwarf star \citep{kirkpatrick}, and its analysis with SPIRou and SOPHIE data led to the discovery of a super-Earth was reported by \citet{CortesZuleta2025}. These authors estimated the separation between these two stars as $63\pm1$ au, with an eccentricity of $0.29\pm0.01$ and an orbital period of $871\pm108$ years. A companion with a mass of 15\,$M_J$ and an orbital period of 5.5 years was reported around Gl\,725\,B by \citet{baize1976}, but this claim was not confirmed by the more recent analysis of \citet{Kervella2019}. A potential planet with a period of 2.7\,d and a $1.2$\,m$\cdot$s$^{-1}$ semi-amplitude was also previously reported around Gl\,725\,B from high-cadence observations with HARPS-N in the optical, but was considered to be comparable to the noise floor \citep{berdinas2016}. 
        
    \subsection{SPIRou data}

    The high-resolution NIR spectropolarimeter SPIRou is installed at the Cassegrain focus of the CFHT in 2018. It covers a spectral range from 0.98 to 2.35\,$\mu$m with a resolving power of 70,000. The instrument was specifically designed for the detection and characterization of exoplanets around low-mass stars in the NIR domain, where M dwarfs are brightest \citep{donati2020}.
    
    \par 
    
    Observations of Gl\,725\,B with SPIRou were conducted between February 14, 2019, and April 19, 2022, resulting in a total of 208 measurements. All spectra were processed with version 0.7.288 of the SPIRou data reduction software \texttt{APERO} \citep{apero}. \texttt{APERO} includes dedicated pipelines for correcting telluric absorption and sky emission lines. It relies on a combination of TAPAS atmospheric models \citep{tapas} and on a library of hot-star observations acquired under varying conditions \citep{artigau2014}. These telluric-corrected spectra form the basis for the subsequent RV extraction.  
    
    \par 
    
    The RVs were obtained using the Line-by-Line (LBL) method\footnote{\url{https://lbl.exoplanets.ca/}}, version 0.65.003. This technique, initially developed by \citet{dumusquelbl} for optical data and extended to the NIR by \citet{etiennelbl}, extracts RVs by tracking the displacement of individual spectral lines. This offers an improved sensitivity for M dwarfs compared to classical cross-correlation techniques \citep{etiennelbl}. It also generates time series for each spectral line individually, providing powerful diagnostics to isolate and characterize systematic effects.
    
    \par 
    
    Instrumental systematics were mitigated in two steps. First, a Fabry-Pérot (FP) etalon, used as a simultaneous reference, allowed us to track short-term instrumental drifts during the night; the measured drift was subtracted from the raw RVs to correct for intra-night velocity variations. Second, a nightly zero point (NZP) correction was applied to account for long-term instrumental drifts, such as those arising from temperature variations or aging of optical fibers. These drifts were monitored using observations of long-term stable stars, and the corresponding NZP correction was subtracted from the nightly RVs.
    
    \par 
    
    Three outliers were identified and removed using a median absolute deviation (MAD) criterion. After these corrections, the dataset achieved a median signal-to-noise ratio per 2.28\,km$\cdot$s$^{-1}$ pixel bin at the center of the H band of 161.9, with a median airmass of 1.355.
    
    \par 
    
    The individual per-line RVs obtained by the LBL method were then analyzed using a weighted principal component analysis (wPCA; \citealt{delchambre}). This technique was recently applied to SPIRou \citep{wapiti} and HARPS \citep{yarara2} data and identifies and removes systematics such as residual telluric contamination. In the next section, we use simulated datasets based on Gl\,725\,B to assess the effect of residual tellurics on the RV precision and to demonstrate the efficiency of the \texttt{wapiti} correction method.

    \subsection{TESS data}
    
    The Transiting Exoplanet Survey Satellite (TESS; \citealt{Ricker2014}) has observed nearly the complete sky in its search for exoplanets around bright stars. Its favorable position in the sky allowed it to observe the binary system Gl\,725 in Sectors 14 to 26, 40, 41, 47 to 57, 59 and 60, covering a period between July 2018 and January 2023. The data from Sector 14 were not considered in our analysis because the position of the star was too close to the edges of the detector. We obtained the presearch data-conditioning (PDC) flux processed by the TESS Science Processing Operations Center (SPOC). The aperture used to perform the flux extraction was not consistent across the sectors, however, meaning that in some sectors, the aperture does not include the two system components. Because the binary components lie so closely to each other and their magnitudes are so similar, it is not possible to independently extract their fluxes. For the sake of consistency, we performed a custom aperture photometry extraction with an aperture that included the fluxes of Gl 725 A and B. 
    
    \par 
    
    The resulting flux time series underwent a cleaning process that removed 3$\sigma$ outliers and a detrending using the \texttt{w$\bar{o}$tan}\footnote{\url{https://github.com/hippke/wotan}} Python package \citep{wotan}, which includes a method based on a time-windowed sliding filter with an iterative robust location estimator. The edges of the TESS time series are usually affected by strong systematics, and we therefore did not consider them in the detrending model. The standard deviation of the detrended light curves is 130 ppm.

    \section{Simulating the effect of telluric absorption lines on LBL data}\label{sec:simulation}
    
    \subsection{Outline of the simulation method}
    
    The limited BERV coverage of Gl\,725\,B  (BERVMAX = 4.6\,km$\cdot$s$^{-1}$) means that telluric contamination leaves systematic residuals in its RV time series, even after correction. To evaluate and quantify this effect and to assess the efficiency of correction methods, we performed dedicated simulations centered on Gl\,725\,B.
    
    \par 
    
    Our approach required an accurate stellar template with minimum telluric residuals. We therefore used the M5.0V dwarf Gl\,905 (HH And, Ross 248; \citealt{gl905type}), which exhibits a large BERV excursion (BERVMAX = 23\,km$\cdot$s$^{-1}$). Its extensive BERV range and high data quality make it suitable for constructing a clean stellar template $T_\star$ with negligible residual telluric contamination.

    \par 
    
    At each epoch, we replaced the observed Gl\,725\,B spectrum with the Gl\,905 template and added photon noise for that epoch. To simulate the effect of telluric contamination, we created a 2D (wavelength times order) telluric template $T_\oplus$ from the output of \texttt{APERO} by taking the median (in the Earth frame) of the telluric spectra derived from Gl\,905 observations \citep{apero}.
    
    \par 
    
    To simulate realistic telluric contamination, we multiplied the noisy stellar template with $T_\oplus$, modulated by a parameter $\alpha$ that encodes imperfect telluric correction,
    
    \begin{equation}\label{eq:simulation_eq}
        T_{contaminated} = T(t) \times T_\oplus^{\alpha(t)}.
    \end{equation}
    
    The $\alpha$ parameter was drawn at each epoch from a normal distribution $\mathcal{N}(\mu_\alpha, \sigma_\alpha)$, with $\mu_\alpha = 3\%$ and $\sigma_\alpha = 1\%$, to reflect typical residual levels after telluric correction \citep{ulmer} and negative values of $\alpha$ mimic overcorrection.
    
    \par 
    
    We considered two cases: (i) the actual BERV values of Gl\,725\,B (narrow coverage), and (ii) scaling the BERV values to match those of Gl\,905 (wide coverage). This allowed us to isolate the specific effect of the BERV excursion on telluric-induced RV signals. 
    
    \subsection{Simulation results}
    
    We generated the RV time series using the LBL algorithm. To ensure realism, we did not reuse the original mask and template from Gl\,905, but recomputed them from the telluric-contaminated spectrum $T(t)$.
    
    \par 
    
\begin{figure*}
        \centering
        \includegraphics[width=\linewidth]{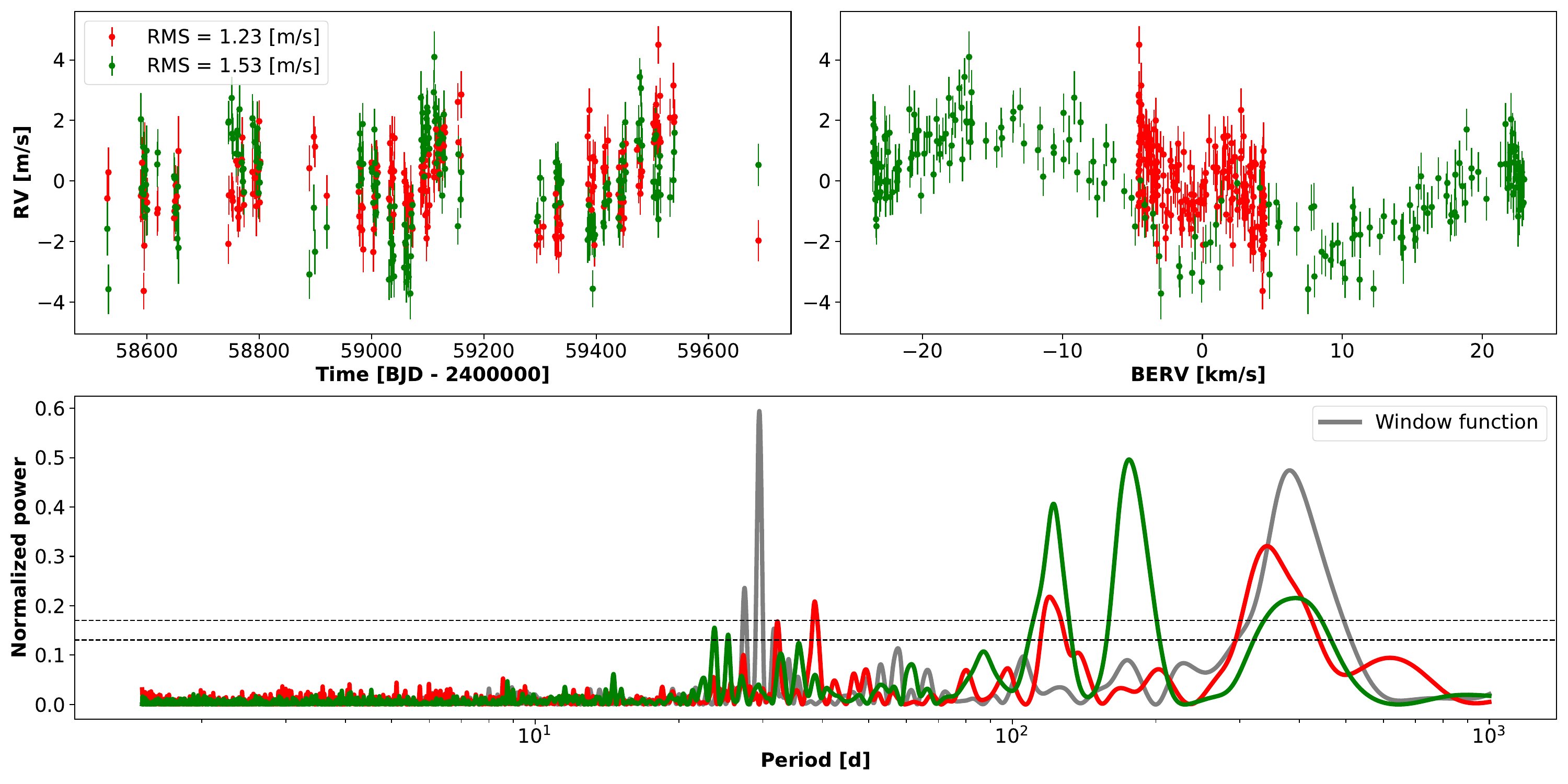}
     \caption[Simulation results when telluric residuals were injected at a $3 \pm 1$\% level]{Simulation results when telluric residuals were injected at a $3 \pm 1$\% level. \textit{Top}: Resulting RV time series from the simulations in time (\textit{left}) and BERV (\textit{right}) spaces. \textit{Bottom}: Their respective GLS periodograms. The RVs computed with a wide and narrow BERV coverage are indicated in red and green, respectively. The significance levels of FAP = $10^{-3}$ and FAP = $10^{-5}$ are indicated by the solid and dashed horizontal lines, respectively. The gray line in the periodogram corresponds to the window function.}
     \label{fig:simulation_results}
    \end{figure*}

    For each BERV scenario (narrow and wide), we ran three independent simulations. As the differences between simulations were minor, we show results from only one simulation per case in Fig.~\ref{fig:simulation_results}, along with RVs as a function of time and BERV. We computed GLS periodograms \citep{Zechmeister_2009} with 10,000 frequency steps between 1.1 and 1,000 days, using the \texttt{astropy} package \citep{astropy}. We determined the false-alarm probability (FAP) of the highest peak following \citet{baluev}.
    
    \par 
    
    The wide and narrow BERV cases show periodic structures at telluric harmonics, with RMS values of 1.23 and 1.53\,m$\cdot$s$^{-1}$, respectively. This agrees with earlier simulations by \citet{tell2}. As expected, even low residual telluric contamination at the percent level induces spurious RV signals. Importantly, this remains true even with wide BERV coverage, which highlights the need for advanced correction methods. Although the RMS is slightly lower in the narrow BERV case, this results from our choice to inject an identical percentage of telluric contamination in both scenarios. This might be misleading because it assumes a relatively optimistic situation for the narrow BERV case: that $\sim$3\% level residuals remain after correction. This is less realistic than for the wide BERV case, where the wider coverage makes low residuals like this more achievable. We also note that telluric contamination affects other LBL observables, such as the differential line width (dLW) that we discuss in Appendix~\ref{section:contamination_stellar_activity}.
    
    \section{Correcting the effect of telluric absorption lines on LBL data}\label{sec:wapiti}
    
    \subsection{The \texttt{wapiti} method}
    
    As discussed in the previous section, Gl\,725\,B presents an interesting challenge through its limited BERV coverage, which increases the effect of residual telluric contamination on the RV measurements. To address this, we tested whether the \texttt{wapiti} method \citep{OuldElhkim2025}, a data-driven approach based on wPCA, can correct for these systematics in the specific context of Gl\,725\,B. A user guide for \texttt{wapiti} is available online\footnote{\url{https://github.com/HkmMerwan/wapiti}}.
    
    \par 
    
    In summary, the \texttt{wapiti} method consists of performing a wPCA on the per-line RVs and selecting the most relevant principal vectors $\hat{V_i}$ to serve as proxy time series to model and correct systematic effects in the RV data. This selection procedure is essential, rather than simply choosing an arbitrary number of leading components, because wPCA captures overall variability in the data and not specifically the systematics. As a result, the first principal component might explain the strongest variance, but not necessarily the dominant systematic signal we aim to correct for. A step-by-step description of this selection process is provided in Appendix \ref{section:wapiti_step}.
    .
    \par 
    
    The components selected for the narrow BERV coverage were the {first, second and tenth components}, while for the wide BERV coverage, they were the {first, third, and fourth components}. One caveat of data-driven methods such as \texttt{wapiti} is that the selected components are not always straightforward to interpret. Ideally, we would understand the effect of telluric absorption lines without relying on wPCA-derived proxies for correction. Nonetheless, in Appendix \ref{section:principal_vector}, we provide an interpretation of the first component, which is common to both cases.
    
    \subsection{Injection-recovery test} 
    
    To evaluate the performance of \texttt{wapiti} for signal detection in RV data affected by tellurics, we carried out an injection-recovery test on the same simulated time series as constructed for Gl\,725\,B. We injected planetary-like signals into the per-line RVs using the form
    
    \begin{equation}\label{signal}
        K\sin\left(\frac{2\pi t}{P} + \phi\right),
    \end{equation}
    
    \noindent where $K$ is the semi-amplitude, $P$ the orbital period, and $\phi$ the phase. We explored a grid of 100 periods logarithmically spaced between 2 and 800 days and 100 semi-amplitudes linearly spaced between 0.5 and 5\,m$\cdot$s$^{-1}$. For each $(P, K)$ pair, we injected the signal ten times using different phases linearly sampled between 0 and $2\pi$.
    
    \par 
    
    Detection was performed using a Bayesian periodogram \citep{Delisle2018}, which compares a null model $\mu_0$ and a planet model $\mu_P$ for each trial period,
    
    \begin{equation}\label{eq:model_0}
       RV(t) = \gamma_{SPIRou} + \sum_{i=1}^{N}a_i\hat{V_i},
    \end{equation}
    \begin{equation}\label{eq:model_p}
        RV(t) = \gamma_{SPIRou} + \sum_{i=1}^{N}a_i \hat{V_i} + A \sin\left(2\pi t / P\right) + B \sin\left(2\pi t / P\right).
    \end{equation}
    
    We computed the Bayesian information criterion (BIC) for both models and derived the logarithm of the Bayes factor as
    \[
    \log BF(P) = \frac{BIC_0 - BIC(P)}{2}.
    \]
    We used a grid of 1,000 points ranging from 1.5\,d to 1,000\,d with a logarithmic spacing, and a threshold of $\log BF > 5$ was used to claim a detection \citep{Delisle2018}.
    
    \par 
    
\begin{figure*}[htbp]
        \centering
        \includegraphics[width=\linewidth]{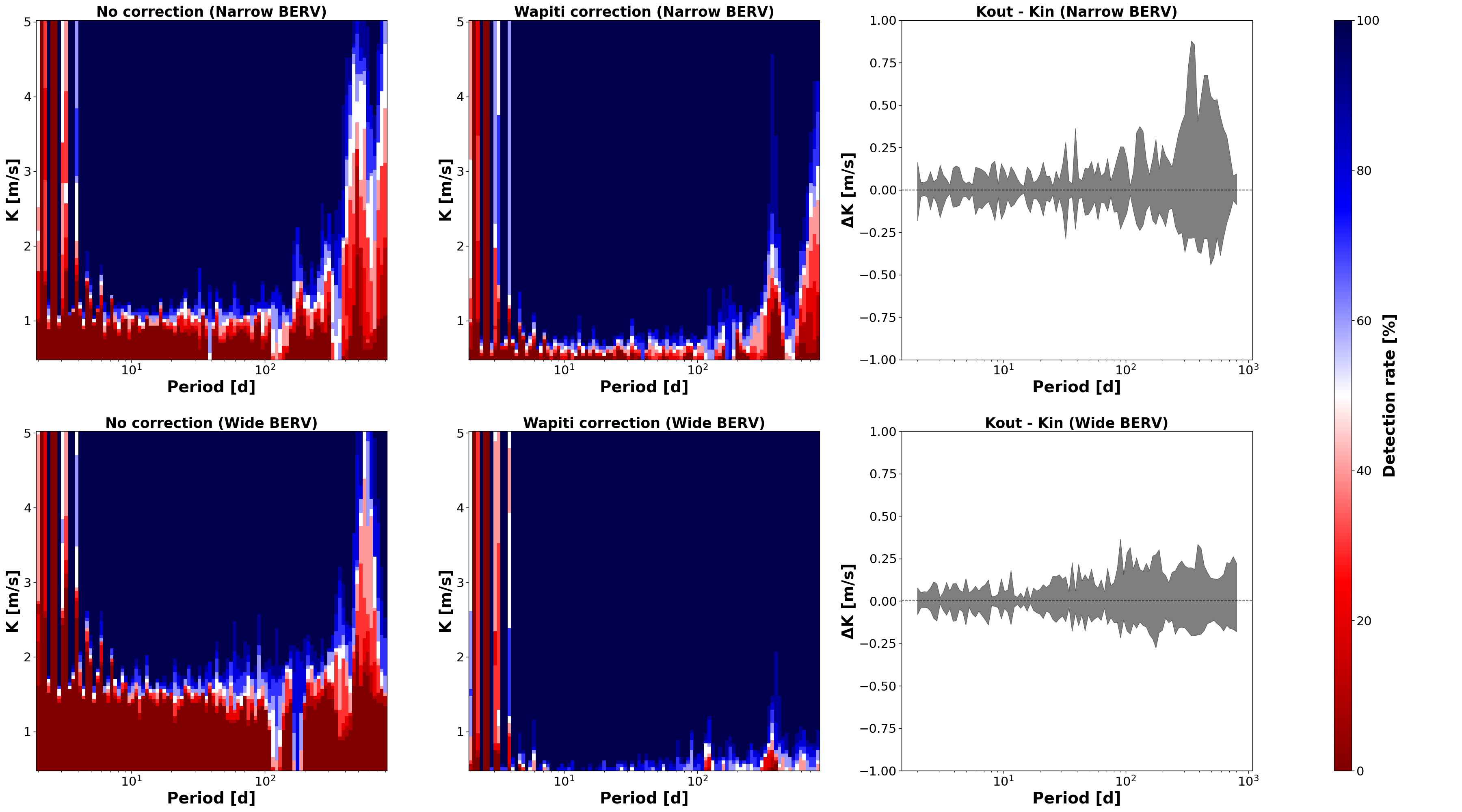}
     \caption[Results of the injection-recovery test] {\textit{Left}: Detection maps showing the percentage of recovered injected signals as a function of period and semi-amplitude for the narrow and wide BERV coverages. \textit{Middle}: Same detection tests after applying the \texttt{wapiti} correction. \textit{Right}: Difference between the recovered and injected $K$ values ($\Delta K = K_\text{out} - K_\text{in}$) at 1\,m$\cdot$s$^{-1}$, with the shaded area indicating one standard deviation. The color bar indicates the detection rate in percent.}
    \label{fig:injection_recovery}
    \end{figure*}

    We performed the test for both BERV scenarios (narrow and wide), with and without the \texttt{wapiti} correction. A signal was considered recovered when the maximum $\log BF$ exceeded 5 \citep{Delisle2018} and was within 5\% of the injected period. The resulting detection maps are shown in Fig.~\ref{fig:injection_recovery}. The detection rates clearly improve after correction, particularly in the low semi-amplitude regime. The signals near one-year aliases and at long periods remain more difficult to recover in the narrow BERV scenario, however.
    
    \par 
    
    To assess whether the semi-amplitude was also correctly retrieved, we tested recovery at $K_\text{in} = 1$\,m$\cdot$s$^{-1}$ and compared it to the recovered value $K_\text{out}$. Fig.~\ref{fig:injection_recovery} shows that \texttt{wapiti} achieved an accurate amplitude recovery within the uncertainties.
    
    \par 
    
    In summary, the \texttt{wapiti} method appears to be effective in correcting systematic effects caused by telluric absorption lines. It improves the detectability of planetary signals and allows for an accurate characterization of their semi-amplitude.
    
    \section{Analysis of Gl\,725\,B}\label{sec:725b}

    \subsection{Activity indicators}
    
    We investigated the stellar activity of Gl\,725\,B using three SPIRou-derived indicators: the dLW, the longitudinal magnetic field ($B_\ell$), and the differential effective temperature (dET; \citealt{dtemp}). The dET indicator is a recently developed proxy for stellar activity that was derived from the LBL analysis of spectral line depth modulations in SPIRou spectra. 
    \par
    
\begin{figure*}[htbp]
        \centering
        \includegraphics[width=\linewidth]{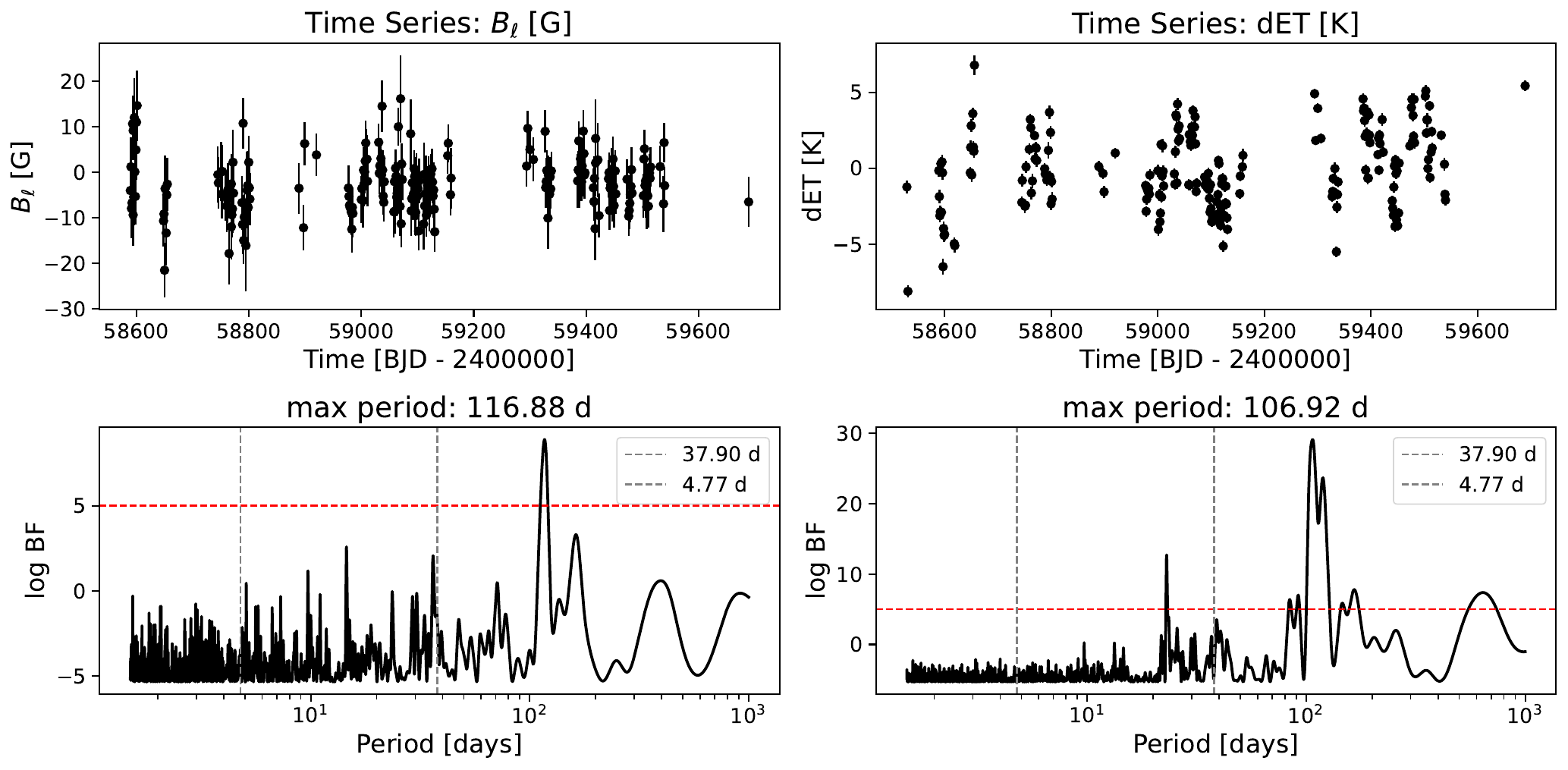  }
     \caption{Time series (\textit{top}) and Bayesian periodograms (\textit{bottom}) for the longitudinal magnetic field $B_\ell$ (\textit{left}) and the differential effective temperature activity indicator dET (\textit{right}). The dashed red line shows the significance threshold of 5. The vertical dashed lines mark the periods of the two signals in RV data at 37.9 and 4.77 days.} 
    \label{fig:activity_indicators}
    \end{figure*}

    The corresponding time series and Bayesian periodograms for $B_\ell$ and dET are shown in Fig. \ref{fig:activity_indicators}, and both reveal strong signals near 117 and 107 days, respectively. These periods are close to the stellar rotation period of $135 \pm 15$ days reported by \citet{donatibell}. The dLW periodogram is heavily contaminated by signals of telluric origins due to the limited BERV coverage of the target (see Appendix \ref{section:contamination_stellar_activity}), which hampers the reliability of this indicator for this source.
    
    \subsection{Correction for stellar activity}
    
    We used the multidimensional Gaussian process (GP) framework proposed by \citet{Rajpaul2015}, which was inspired by the work of \citet{aigrain2012}. It enables the simultaneous modeling of stellar activity and Keplerian signals in the RV time series by incorporating information from stellar activity indicators.
    
    \par 
    
    In this framework, the stellar activity affecting the RVs and activity indicators is described by a common latent function ${GP(t)}$ and its time derivative $\dot{GP}(t)$. The activity-induced RV variations are then expressed as linear combinations of ${GP(t)}$ and $\dot{GP}(t)$. For this analysis, we employed the \texttt{s+leaf 2} GP framework\footnote{\url{https://dace.unige.ch/pythonAPI/?tutorialId=703}}, which is designed for an efficient joint modeling of multiple time series with a computational cost that scales linearly with the dataset size \citep{Delisle2022}. We adopted a Matérn 3/2 exponential periodic (MEP) kernel, which approximates the squared-exponential periodic (SEP) kernel that was widely used to constrain stellar activity signals \citep{aigrain2012, Haywood_2014},
    
    \begin{equation}
        k\left(\tau\right) = \sigma \exp\left(- \frac{\tau^2}{2\rho^2} - \frac{\sin^2\left(\frac{\pi}{P}\tau \right)}{2\eta^2}\right).
        \label{gp_form}
    \end{equation}
    
    We tested two multidimensional GP models: one model including the RV and dET time series (RV + dET), and another model with the RV, dET, and $B_\ell$ time series (RV + dET +  $B_\ell$). The models are described as follows:
    
\begin{equation}\label{multiGP}
\begin{aligned}
RV      &= M_0 + \alpha_1\,GP(t) + \beta_1\,\dot{GP}(t), \\
dET     &= \mathrm{offset}_{dET} + \alpha_2\,GP(t) + \beta_2\,\dot{GP}(t), \\
B_\ell  &= \mathrm{offset}_{B_\ell} + \alpha_3\,GP(t) + \beta_3\,\dot{GP}(t), \\[3pt]
M_{0}   &= \gamma + \mathrm{trend}\cdot t 
          + \sum_{i=1}^{N} a_i\,\hat{V}_i 
          + \sum_{i=1}^{N_p} \left[ 
              A_i\,\sin\!\left(\tfrac{2\pi t}{P_i}\right)
              + B_i\,\cos\!\left(\tfrac{2\pi t}{P_i}\right)
            \right],
\end{aligned}
\end{equation}

    \noindent where $M_0$ is the base model incorporating a constant offset $\gamma$, a trend term, systematic components $\hat{V_i}$, and previously identified planetary signals at periods $P_i$ (if any).
    
    To assess the importance of the derivative term  $\dot{GP(t)}$, we explored all combinations of setting the $\beta$ coefficients to zero or allowing them to vary, using likelihood maximization to fit the models. The results (Table \ref{tab:model_comparison_gl725B}) show that in the RV + dET model, fixing both $\beta_1$ and $\beta_2$ to zero is favored. Physically, ${GP(t)}$ is interpreted as related to the area of the visible stellar disk covered by active regions, while its time derivative corresponds to the temporal evolution of these regions. Therefore, the lack of preference for including the derivative in the RV and dET model suggests that convective blueshift suppression is the dominant activity-induced effect in these time series \citep{Dumusque2014}. In contrast, for the RV + dET + $B_\ell$ model, the favored configuration also sets $\beta_1$ and $\beta_2$ to zero, but allows $\beta_3$ to vary freely. This indicates that for the $B_\ell$ time series, the evolution of active regions on the stellar surface plays a more significant role in the observed activity signal.

    \subsection{Search for signals in the RVs}
    
    {To search for signals in the RV time series, we employed an iterative approach based on Bayesian periodograms \citep{Delisle2018, OuldElhkim2025}. In this method, significant signals are sequentially added to the model, and the log BF periodogram is recomputed at each step. The GP hyperparameters were initially taken from the best fit of the multidimensional model described by Eq. \ref{multiGP}. During the period search, all GP hyperparameters were kept fixed except for the amplitude $\alpha$, while we computed the log BF over a dense grid of 10,000 trial periods between 1.5 and 1,000 days by adding a circular planetary signal at each fixed period.}

    {When a significant signal at period $P$ was detected during iteration $N_{p+1}$, a new model $M_P$ was tested by adding a sinusoidal component at that period. Because this search assumes fixed periods and GP hyperparameters, it can slightly overestimate log BF. After each detection, we therefore refit the full model to allow the orbital period and all GP hyperparameters to vary freely in order to verify the statistical robustness of the signal}
    
    \par 
    
\begin{figure*}[htbp]
        \centering
        \includegraphics[width=\linewidth]{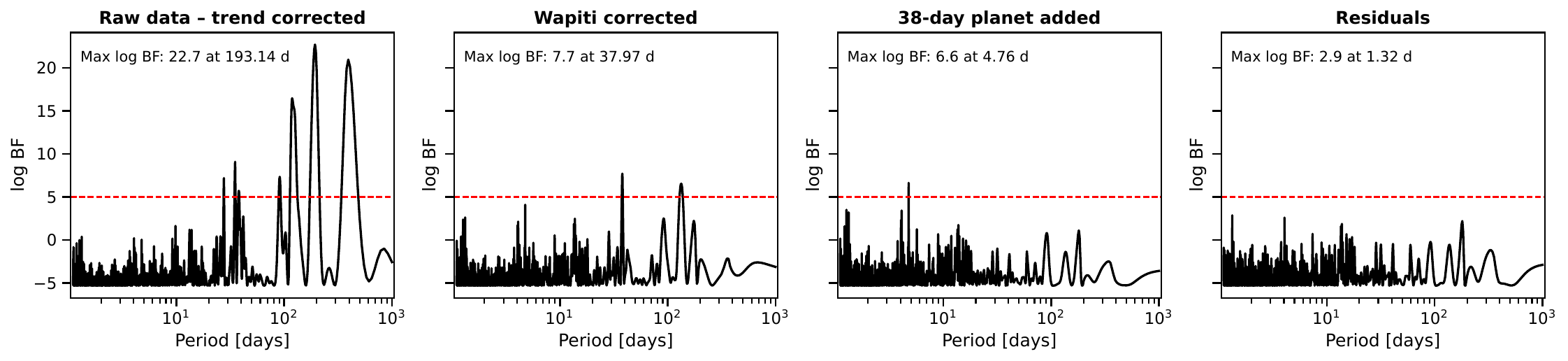}
     \caption{Iterative Bayesian periodograms for the Gl\,725\,B RV time series under different models. From left to right: (1) Raw RVs after trend correction, (2) after removal of systematics via wapiti, (3) with the 38-day planet, and (4) residuals after modeling the 38-day and 4.77-day signals. The dashed red line marks the detection threshold at log BF = 5.} 
    \label{fig:bayesian_periodograms}
    \end{figure*}

\begin{figure*}[htbp]
    \centering
    \includegraphics[width=\linewidth]{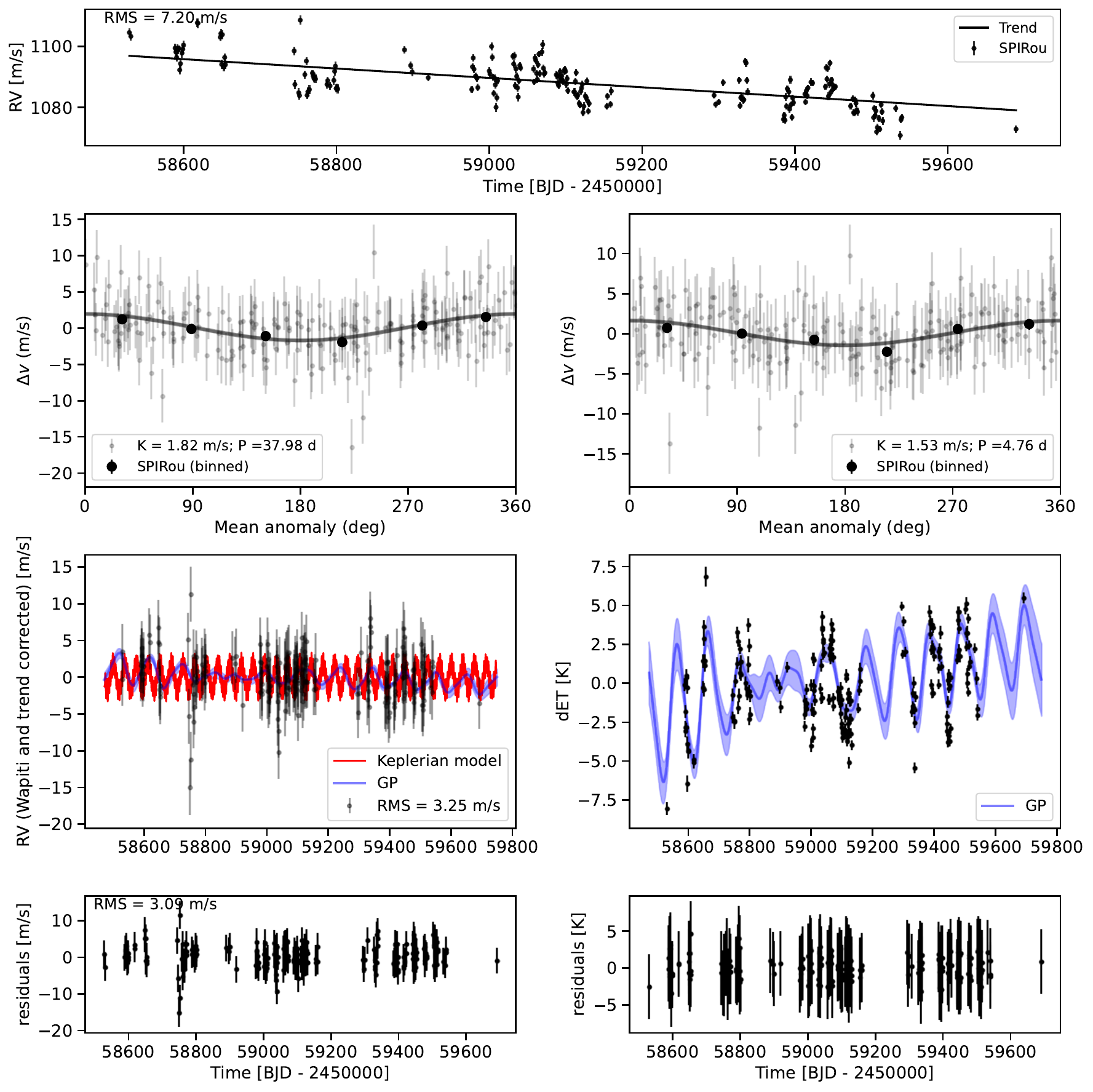}
    \caption{Two-planet GP model for Gl\,725\,B. \textit{Top}: Raw RVs with trend. \textit{Middle}: Phase-folded RVs for the 38-day and 4.77-day signals with binned points (black). \textit{Bottom}: Modeled RV and dET time series (with GP) and corresponding residuals.}
    \label{fig:model_steps}
    \end{figure*}

    The \texttt{wapiti} method applied on Gl\,725\,B dataset selected the second, ninth, first, and seventh components (see Appendix \ref{section:wapiti_step} for a detailed explanation of this component selection), which successfully removed systematic signals at periods of 180 days, 1 year, and 120 days. The iterative Bayesian periodograms are shown in Fig. \ref{fig:bayesian_periodograms}. After applying the wapiti correction, we detected a 38-day signal, and when this was added to the model, another 4.77\,d signal was detected in the residuals. Finally, after adding this second signal, {no further signals were detected}.     {We found no trace of the 2.7\,d signal reported by \citet{berdinas2016}, nor evidence of the 5.5-year companion of \citet{baize1976}. Based on their parameters, if such a companion existed,  it would produce a semi-amplitude of about 600\,m$\cdot$s$^{-1}$. This amplitude is clearly absent from our SPIRou RVs (Fig. \ref{fig:model_steps}).}
    
    \par 
    
    Consequently, we fit a one-planet (1p) model including the 38-day signal and a two-planet (2p) model including the 38-day and 4.77-day signals, considering both circular and Keplerian orbits, by maximizing their likelihood. The results are presented in Table \ref{tab:model_comparison_gl725B}. {The RV + dET model yielded the best result, with log BF of 4.1 and 6.2 for the one-planet and two-planet cases, respectively, compared to the model without planetary signals (see Table \ref{tab:model_comparison_gl725B}).}
    
    {These log BF values were computed from the maximum likelihood values alone and might misrepresent the true evidence \citep{Perrakis_2014}, however. A more rigorous assessment of the signal significance, for example, using the false-inclusion probability (FIP; \citealt{fip}) periodogram, which makes use of the full posterior distribution, is presented in Appendix \ref{section:fip}. In this posterior-based analysis, only the 38-day signal is robustly recovered, while the 4.77-day signal does not reach the detection threshold.}

    {Additionally, we further verified the robustness of these periodicities with respect to the choice of the NZP correction. This sensitivity analysis, presented in Appendix~\ref{app:nzp_sensitivity}, confirmed that the 38-day signal was consistently recovered in the NZP realizations, whereas the short-period signal is more sensitive to the specific NZP choice and was therefore considered as a candidate rather than a firm detection.}

    \begin{table*}[htbp]
    \centering
    \caption[]{Comparison of the model performances for Gl\,725\,B using different combinations of time series.}
    \small
    \begin{tabular}{l l c c c c c}
    \hline
    Time Series & Model & $\log \mathcal{L}_{\rm max}$ & $\chi^2_{\rm red}$ & BIC & $\log \mathrm{BF}$ & RMS [m/s] \\
    \hline
    RV only & offset + wapiti + trend & -552.04 & 1.050 & 1141.33 & 0 & 3.6 \\
     & offset + wapiti + trend + 1p (circular) & -540.16 & 1.068 & 1133.55 & 3.9 & 3.4 \\
     & \textbf{offset + wapiti + trend + 2p (circular)} & \textbf{-530.63} & \textbf{1.090} & \textbf{1130.47} & \textbf{5.4} & \textbf{3.3} \\
     & offset + wapiti + trend + 2p (eccentric) & -529.67 & 1.113 & 1149.83 & -4.3 & 3.2 \\
    \hline\hline
    RV + dET & offset + wapiti + trend & -1023.37 & 1.231 & 2100.96 & -36.6 & 3.8 \\
     & offset + wapiti + trend + GP ($\beta_1$ = $\beta_2$ = 0) & -971.71 & 0.768 & 2027.76 & 0 & 3.5 \\
     & offset + wapiti + trend + GP ($\beta_1$ = $\beta_2$ = 0) + 1p (circular) & -958.56 & 0.770 & 2019.52 & 4.1 & 3.3 \\
     & \textbf{offset + wapiti + trend + GP ($\beta_1$ = $\beta_2$ = 0) + 2p (circular)} & \textbf{-947.44} & \textbf{0.751} & \textbf{2015.35} & \textbf{6.2} & \textbf{3.1} \\
     & offset + wapiti + trend + GP ($\beta_1$ = $\beta_2$ = 0) + 2p (eccentric) & -946.39 & 0.755 & 2037.35 & -4.8 & 3.1 \\
    \hline\hline
    RV + dET + $B_\ell$ & offset + wapiti + trend & -1680.02 & 1.754 & 3430.79 & -34.8 & 3.8 \\
    & offset + wapiti + trend + GP  ($\beta_1$ = $\beta_2$ = 0) & -1622.75 & 1.229 & 3361.27 & 0 & 3.5 \\
    & offset + wapiti + trend + GP  ($\beta_1$ = $\beta_2$ = 0) + 1p (circular) & -1609.92 & 1.286 & 3354.89 & 3.2 & 3.3 \\
    & \textbf{offset + wapiti + trend + GP  ($\beta_1$ = $\beta_2$ = 0) + 2p (circular)} & \textbf{-1598.76} & \textbf{1.324} & \textbf{3351.86} & \textbf{4.7} & \textbf{3.1} \\    
    & offset + wapiti + trend + GP  ($\beta_1$ = $\beta_2$ = 0) + 2p (eccentric) & -1597.59 & 1.356 & 3381.69 & -10.2 & 3.1 \\
    \hline
    \end{tabular}
    \begin{minipage}{\textwidth}
    \small
    Maximum log-likelihood, reduced chi-squared, BIC, log BF, and RMS of the residuals for the best model applied to each combination of time series. The model in bold indicates the best-fit model to the combination of time series.
    \end{minipage}
    \label{tab:model_comparison_gl725B}
    \end{table*}    
    
    \subsection{Characterization of the system}
    
    The absence of the two signals in the stellar activity indicators, combined with their robustness and statistical significance even when stellar activity was modeled in the RV data, {supports a planetary interpretation for the 38-day signal and suggests that the 4.77-day signal is a plausible planetary candidate}. To characterize the detected signals, we used the adaptive MCMC sampler samsam\footnote{https://gitlab.unige.ch/Jean-Baptiste.Delisle/samsam} \citep{Delisle2018}. The sampler was configured to perform a total of 1,000,000 iterations, and we used the final 750,000 iterations for the statistical analysis. We ran two cases, one case with circular orbits, and the one with Keplerian orbits. 
    
    \par 
    
    The results are presented in Table \ref{gl725b_orbital_parameters} and show that the eccentricities of {the confirmed planet and the candidate signal} cannot be well constrained with the current dataset. This is expected given the low amplitude of the signals, and it is consistent with Table \ref{tab:model_comparison_gl725B}, where the Keplerian model was not favored over the circular model. Nevertheless, all other orbital parameters remained consistent between the circular and eccentric fits. The circular model therefore likely provides a more reliable representation of the system. The corner plot for this final circular fit is shown in Appendix \ref{section:mcmc_gl725b}.
    
    \begin{table}[htbp]
    \scriptsize
    \setlength{\tabcolsep}{7pt}
     \centering
    \caption{Orbital parameters of Gl\,725\,B, {showing the best-fit model including one confirmed planet (c) and one candidate (b)} with circular and Keplerian fits.}
    \begin{tabular}{|l|cc|c|}
    \hline
    \multirow{2}{*}{Parameters} & \multicolumn{2}{c|}{2p} & \multirow{2}{*}{Priors} \\
    \cline{2-3}
    & Circular & Keplerian & \\
    \hline
    \multicolumn{4}{|c|}{\textbf{Instrumental Parameters}} \\
    \hline
    $\gamma_{SPIRou}$ (m\,s$^{-1}$) & $1143.7 \pm 5.7$ & $1143.9 \pm 5.3$ & $\mathcal{U}(-\infty, \infty)$ \\ 
    trend (m\,s$^{-1}$\,d$^{-1}$) & $-0.016 \pm 0.001$ & $-0.016 \pm 0.001$  & $\mathcal{U}(-\infty, \infty)$ \\ 
    $\sigma_{SPIRou}$ (m\,s$^{-1}$) & $2.9 \pm 0.2$ & $2.9 \pm 0.2$ & $\mathcal{LU}(10^{-4}, 10^4)$ \\ 
    \hline
    \multicolumn{4}{|c|}{\textbf{wapiti Systematics}} \\
    \hline
    $a_1$ (m/s) & $-34.1 \pm 3.4$ & $-34.2 \pm 3.4$ & $\mathcal{U}(-\infty, \infty)$ \\
    $a_2$ (m/s) & $-25.5 \pm 3.3$ & $-25.7 \pm 3.3$ & $\mathcal{U}(-\infty, \infty)$ \\ 
    $a_3$ (m/s) & $-19.7 \pm 3.4$ & $-19.6 \pm 3.4$ & $\mathcal{U}(-\infty, \infty)$ \\ 
    $a_4$ (m/s) & $15.8 \pm 3.5$ & $16.1 \pm 3.5$ & $\mathcal{U}(-\infty, \infty)$ \\ 
    \hline
    \multicolumn{4}{|c|}{\textbf{dET Parameters}} \\
    \hline
    $dET$ (K) & $0.5 \pm 8.9$ & $0.2 \pm 7.8$ & $\mathcal{U}(-\infty, \infty)$ \\
    $\sigma_{dET}$ (K) & $1.6 \pm 0.1$ & $1.6 \pm 0.1$ & $\mathcal{LU}(10^{-4}, 10^4)$ \\
    \hline
    \multicolumn{4}{|c|}{\textbf{Stellar Activity Hyperparameters}} \\
    \hline
    $P_{\mathrm{rot}}$ (days) & $105.1 \pm 3.3$ & $105.1 \pm 3.4$ & $\mathcal{LU}(0.1, 10^5)$ \\
    $\rho$ (days) & $2160 \pm 1697$ & $2072 \pm 1644$ & $\mathcal{LU}(0.1, 10^5)$ \\
    $\eta$ & $1.3 \pm 0.5$ & $1.3 \pm 0.5$ & $\mathcal{LU}(0.05, 20)$ \\
    $\alpha_1$ (m/s) & $-5.3 \pm 2.8$ & $-5.0 \pm 2.6$ & $\mathcal{N}(0, 10)$ \\
    $\alpha_2$ (K) & $10.1 \pm 4.9$ & $9.6 \pm 4.7$ & $\mathcal{N}(0, 10)$ \\
    \hline
    \multicolumn{4}{|c|}{\textbf{Orbital Parameters}} \\
    \hline
    $P_b$ (days) & $4.765 \pm 0.004$ & $4.765 \pm 0.004$ & $\mathcal{LU}(0.1, 10^5)$ \\
    $K_b$ (m\,s$^{-1}$) & $1.4 \pm 0.3$ & $1.4 \pm 0.4$ & $\mathcal{LU}(0.1, 10^4)$ \\
    $\lambda_b(0)$ (deg) & $-0.3 \pm 3.8$ & $-0.3 \pm 3.9$ & $\mathcal{U}(0, 2\pi)$ \\
    $e_b \cos\omega_b$ & -- & $0.00 \pm 0.12$ & $\mathrm{modBeta}(0.867, 3.03)$ \\
    $e_b \sin\omega_b$ & -- & $-0.03 \pm 0.13$ & $\mathrm{modBeta}(0.867, 3.03)$ \\
    $P_c$ (days) & $37.90 \pm 0.17$ & $37.91 \pm 0.16$ & $\mathcal{LU}(0.1, 10^5)$ \\
    $K_c$ (m\,s$^{-1}$) & $1.7 \pm 0.3$ & $1.7 \pm 0.4$ & $\mathcal{LU}(0.1, 10^4)$ \\
    $\lambda_c(0)$ (deg) & $5.2 \pm 2.6$ & $5.4 \pm 2.5$ & $\mathcal{U}(0, 2\pi)$ \\
    $e_c \cos\omega_c$ & -- & $0.05 \pm 0.14$ & $\mathrm{modBeta}(0.867, 3.03)$ \\
    $e_c \sin\omega_c$ & -- & $-0.03 \pm 0.14$ & $\mathrm{modBeta}(0.867, 3.03)$ \\
    \hline
    \multicolumn{4}{|c|}{\textbf{Derived Parameters}} \\
    \hline
    $\mathrm{M_{p,b}}\sin i$ (M$_\oplus$) & $1.5 \pm 0.4$ & $1.4 \pm 0.4$ & -- \\
    $a_b$ (AU) & $0.035 \pm 0.001$ & $0.035 \pm 0.001$ & -- \\
    $e_b$ & -- & $0.13 \pm 0.11$ & $\mathrm{Beta}(0.867, 3.03)$ \\
    $\omega_b$ & -- & $-0.4 \pm 1.8$ & -- \\
    T$_\mathrm{eq,b}$ (K) & $430 \pm 8$ & $430 \pm 8$ & -- \\
    $\mathrm{M_{p,c}}\sin i$ (M$_\oplus$) & $3.4 \pm 0.7$ & $3.5 \pm 0.7$ & -- \\
    $a_c$ (AU) & $0.139 \pm 0.004$ & $0.139 \pm 0.004$ & -- \\
    $e_c$ & -- & $0.14 \pm 0.12$ & $\mathrm{Beta}(0.867, 3.03)$ \\
    $\omega_c$ & -- & $-0.3 \pm 1.5$ & -- \\
    T$_\mathrm{eq,c}$ (K) & $215 \pm 4$ & $215 \pm 4$  & -- \\
    \hline
    \end{tabular}
    \label{gl725b_orbital_parameters}
    \end{table}
    
    Our multidimensional GP analysis yields a stellar rotation period of $P_{rot} = 105.1 \pm 3.3$\,d, which is more precise than the previously reported value of $135 \pm 15$\,d based on the $B_\ell$ time series \citep{donatibell}. The difference in value between these two estimates (but they are still compatible within 2\,$\sigma$) might arise from our earlier finding that for the $B_\ell$ time series, the evolution of active regions contributes more significantly to the observed activity signal, whereas this effect appears to be less prominent in the RV and dET time series.
    
    \par 
    
    As for the planetary signals, the minimum mass of these planets increases inside-out, ranging from $1.5 \pm 0.4$\,$M_\oplus$ for {the candidate} Gl\,725\,Bb to $3.4 \pm 0.7$\,$M_\oplus$ for Gl\,725\,Bc. According to the classification by \citet{mishra2023}, the architecture of this system falls into the "ordered" class. This class was identified as relatively rare in simulations, with a 1.5\% occurrence rate. In contrast, their observation sample indicates a higher prevalence, with 37\% of observed systems falling into this class. Therefore, our findings support the idea that ordered planetary systems may be more common than predicted by simulations. Based on \citet{hz} and using the stellar parameters from table \ref{gl725b_orbital_parameters}, the habitable zone (HZ) of Gl\,725\,B for an Earth-mass planet spans from 0.09 to 0.19 AU. Consequently, Gl\,725\,Bc falls within the habitable zone of its host star. We discuss the localization of Gl\,725\,Bc in the habitable zone in more detail in Sect. \ref{sec:planetary_system}.
    
    \par 
    
    We did not have to take the effect of the binary companion Gl\,725\,A into account because both planets lie at a distance that is smaller than the critical semimajor axis of the system, $a_c = 12.4$\,AU. This value was calculated using the first formula from \citet{holman1999} and the orbital parameters from \citet{CortesZuleta2025} for the binary system. Using these parameters, we also calculated an expected trend of $0.0238 \pm 0.0059$\,m$\cdot$s$^{-1}$.d$^{-1}$, which is consistent within uncertainties with the trend obtained from the MCMC analysis of $0.016 \pm 0.001$\,m$\cdot$s$^{-1}$.d$^{-1}$.

    \subsection{Analysis of the TESS photometric data}

    We used the \texttt{transit least squares} algorithm\footnote{\url{https://github.com/hippke/tls}} (TLS; \citealt{tls}) to search for transit events in the blended light curves of Gl 725 A and B. No transits were identified by the TLS, including periodic events at the orbital period of planet b. Because of the long time span of the TESS data, a periodic event of 4 days or even longer than 30 days is expected to be easy to identify. The derived planetary parameters suggest a transit probability of only 3.7$\%$ for planet b and 0.9$\%$ for planet c, however, which might explain the no-transiting nature of the two planets.

    To assess the transit detection sensitivity, we conducted an injection-recovery test in our TESS data for a range of orbital inclinations and planet radii following \citet{CortesZuleta2025}. For each planet, we performed 50\,000 simulations of transit events generated by the \texttt{batman} Python package \citep{batman}. The planet parameters were taken from Table \ref{gl725b_orbital_parameters}, and the limb-darkening coefficients were taken from Table 15 in \citet{Claret2017}. For each run, the orbital inclination and planet radius were drawn randomly from uniform distributions between 85$^{\circ}$ and 90$^{\circ}$ for the inclination and 0.5 to 10$R_\oplus$ for the radius. Then, we applied the \texttt{box least squares} (BLS, \citealt{bls}) periodogram from Astropy\footnote{\url{https://docs.astropy.org/en/stable/timeseries/bls.html}} to measure the power of the signal at the expected orbital period of planet b (4.765 days) and planet c (37.90 days).
    
\begin{figure}[htbp]
        \centering
        \includegraphics[width=\linewidth]{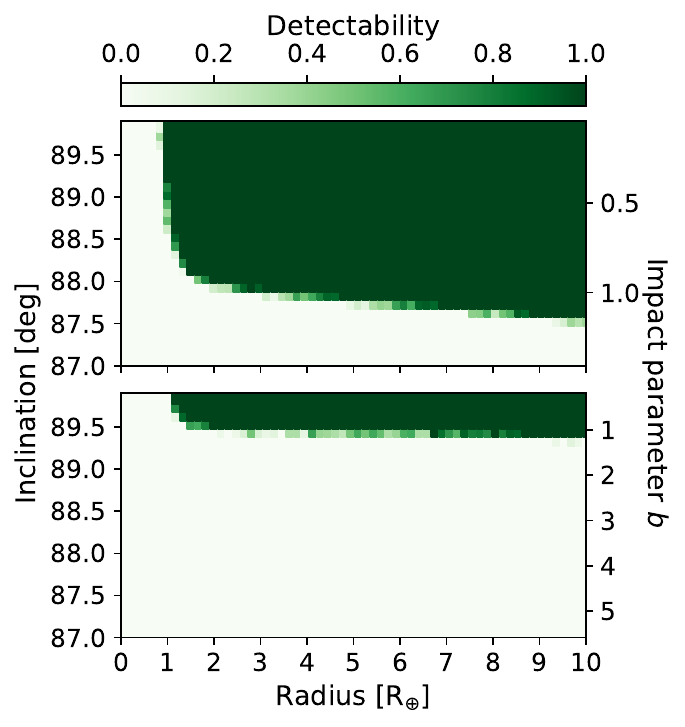}
    \includegraphics[width=\linewidth, trim=0 0 0 10, clip]{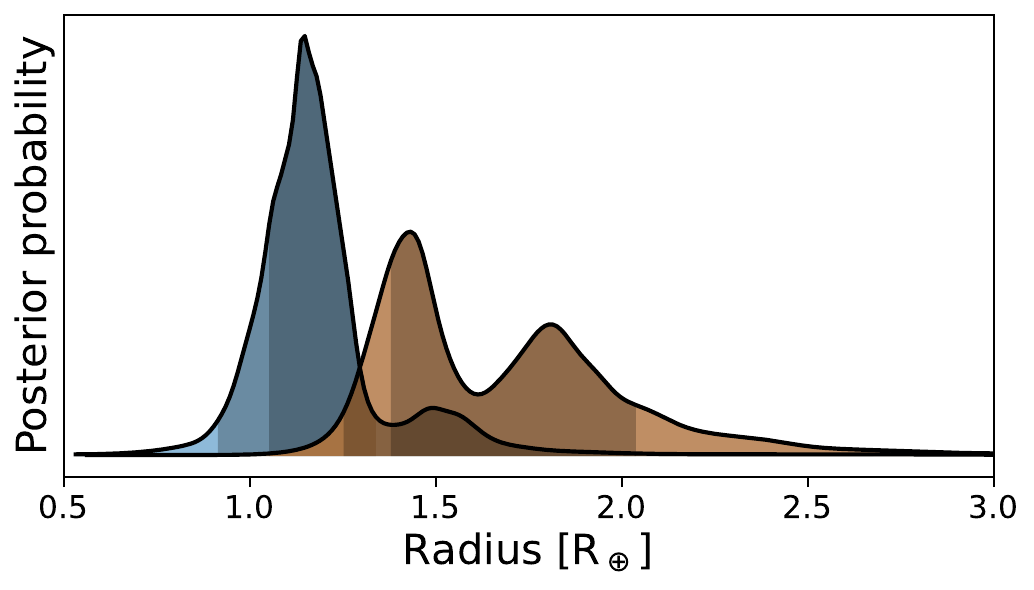}
    \caption{Detectability of transit events of {the candidate} planet b (\textit{top}) and {the confirmed} planet c (\textit{middle}) in the TESS data of Gl 725 AB as a function of orbital inclination and planet radius. \textit{Bottom}: Posterior distributions of the planetary radii for Gl\,725\,Bb (blue) and Gl\,725\,Bc (brown) inferred using the \texttt{spright} tool \citep{spright}.}
    \label{fig:gl725B_injection}
    \end{figure}

    We binned the orbital inclination and planet radius in bins of 0.1 units for each parameter and then computed the fraction of transit events that are detectable with the BLS periodogram (i.e., power > 2000) in each bin. Since the light curves are blended, the true detectable planet radius needs to be scaled by the dilution factor, which has a value of 1.64. The results are shown in Fig. \ref{fig:gl725B_injection}, where we found that in general, the detection limit of the TESS photometry is about 1.0$R_\oplus$ for aligned planets. For planet b, most of the planets with a radius greater than 1.0$R_\oplus$ and orbital inclinations between 87.5$^{\circ}$ and 90$^{\circ}$ could be detected. In the case of planet c, the transit would have been detected for orbital inclinations between 89.4$^{\circ}$ and 90$^{\circ}$ and for a planet radius larger than 1.2$R_\oplus$. For both cases, smaller planets are detectable for inclinations closer to 90$^{\circ}$, as expected.
    
    \par 
    
    {Using the \texttt{spright} tool \citep{spright}, we predict} the radius of Gl\,725\,Bb to be \(1.2^{+0.2}_{-0.1}\,R_\oplus\) and that of Gl\,725\,Bc to be \(1.7^{+0.4}_{-0.3}\,R_\oplus\) (see Fig. \ref{fig:gl725B_injection}). The posterior distribution for planet c is bimodal, which is known as the "radius valley" \citep{fulton2017, cloutier2020, gaidos2024}, and it was approximated here as the sum of two Gaussian components centered at $1.41 \pm 0.09$\,R$_\oplus$ and $1.82 \pm 0.20$\,R$_\oplus$. The transit events of planet~b are undetectable if the planet radius is $\leq1.0\,R_\oplus$, which is unlikely from the posterior distribution of predicted planet radii of Fig. \ref{fig:gl725B_injection}. Similarly, planet~c is unlikely to be transiting because we predict a planet radius larger than \(1.2\,R_\oplus\), which is detectable in the TESS photometry.

    \par 
    
    \section{Discussion and conclusions}\label{sec:ccl}
    
    \subsection{Addressing the RV effects of telluric absorption lines in the NIR using wapiti}
    
    To understand and mitigate the effect of telluric absorption lines on RV measurements for Gl\,725\,B, we performed controlled simulations by injecting realistic telluric residuals at the $3 \pm 1$\% level into a clean stellar template. These simulations were designed to replicate the BERV constraints of Gl\,725\,B and showed that even modest telluric residuals induce systematic signals in the RVs extracted with the LBL algorithm. This reinforces the need for dedicated correction methods, such as \texttt{wapiti}, especially for targets with limited BERV coverage.
    
    \par 
    
    Using the per-line RV time series from our simulations, we conducted injection-recovery tests and found that \texttt{wapiti} successfully corrected for the effects of telluric contamination. In the narrow and wide BERV scenarios, it improved the detectability and enabled an accurate recovery of the planetary signal amplitudes.
    
    \par 
    
    While this is promising, it is important to recognize the limitations of these simulations. They represent a simplified environment in which the star is modeled as a static template plus photon noise. In reality, additional complexities arise, including seasonal or chromatic variability in telluric residuals ($\alpha$), night-sky emission lines (OH and Na), Moon contamination \citep{crepp}, and variations in the blaze function due to instrumental or illumination changes \citep{halverson}. These are not captured in our controlled setup, but would affect real observations.
    
    \par
    
    In addition, a core limitation of \texttt{wapiti} (or any wPCA-based methods) is that wPCA is designed to capture variability rather than systematics. Although the \texttt{wapiti} method is implemented in a way that accounts for this limitation, it remains a key aspect to consider because many of its capabilities and limitations arise from it. For example, because of this tendency to capture variability, wPCA is highly sensitive to outliers and observations with a low signal-to-noise ratio, and it might be necessary to exclude some measurements in certain situations.
    
    \par 
    
    Finally, since wPCA constructs orthogonal components by design, real-world systematics that are not orthogonal might become entangled within the same vector. This would complicate the interpretation. Despite these limitations, \texttt{wapiti} remains a powerful tool for mitigating telluric-induced variability, as we demonstrate below in the analysis of real data from Gl\,725\,B.

    \subsection{Stellar age}
    \label{sec:age}
    
    Our multidimensional GP analysis retrieved a rotational period of $105.1 \pm 3.3$\,d for Gl\,725\,B. We combined this value with the $T_{\rm eff}$ derived by \citet{cristofari} with the gyrochronology presented by \citet{Gaidos2023} to estimate an age of 8.7 $\pm$ 2.1 Gyr.  This is consistent with the old but very uncertain age of Gl\,725A \citep[$\gtrsim 7$ Gyr,][]{Fouque2023,CortesZuleta2025}.  The correction for a nonsolar metallicity (different from the stars used to calibrate the gyrochronology) was estimated at +0.8 Gyr, but this is only based on theory.  Three other caveats are that (1) the estimated age of Gl\,725B is much older than the oldest calibrator \citep[the 4 Gyr-old M67 cluster][]{Dungee2022}, and Skumanich-like spin-down \citep{Skumanich1972} is assumed after this point,  (2) Gl\,725B is near the cool end of the valid range, and (3) Gl\,725B is in a binary system with a projected separation of 65\,au, where its protostellar disk would have been truncated, likely leading to more rapid dissipation, less angular momentum exchange with the star, and hence, more rapid initial stellar rotation, as was observed in young clusters \citep[e.g.,][]{Kraus2012,Messina2019}.  Nevertheless, that the star and its planets are old is likely a reliable result.  The kinematics of the system ($U = -25.33$,   $V=-12.95$, $W=26.58$ km\,sec$^{-1}$) are consistent with the Galactic thin disk and are not particularly suggestive of nor inconsistent with an old age.

    \subsection{Planetary system Gl\,725\,B}\label{sec:planetary_system}
    
{Through the \texttt{wapiti} correction, we uncovered evidence of a compact ordered planetary system orbiting Gl\,725\,B. 
The outer planet, Gl\,725\,Bc, is firmly detected with a minimum mass of $3.4 \pm 0.7$\,M$_\oplus$ and an orbital period of $37.90 \pm 0.17$ days. 
In addition, we identified a shorter-period signal at $4.765 \pm 0.004$ days that might correspond to a possible inner planet candidate, Gl\,725\,Bb, with a minimum mass of $1.5 \pm 0.4$\,M$_\oplus$.
Further observations are required to confirm its planetary nature.
The system thus appears to consist of at least one confirmed planet (Gl\,725\,Bc) and a promising close-in candidate.}

    \par

    Recent estimates of the luminosity of Gl725B derived from stellar radius measurements, empirical relations, or the fitting of synthetic spectra consistently converged to between $8.4 \times 10^{-3}$ and $9.2 \times 10^{-3}$ L$_\odot$ \citep{cristofari, Boyajian2012, Mann2015}. Based on the semi-major axes of 0.035 AU and 0.139 AU for planets Gl725Bb and Gl725Bc, respectively, their received stellar fluxes are approximately 7 and 0.45 times that of Earth. The latter value places Gl725Bc in an insolation regime analogous to that of Mars, and it positions it securely within the nominal circumstellar habitable zone.
    \par
    
    Significant recent progress has been made in characterizing the inner boundary of the habitable zone, particularly in relation to the onset of the runaway greenhouse effect. This has been achieved using both 1D climate models \citep{Kasting1993, hz, Yang2016, Chaverot2022} and 3D global climate models \citep{Wordsworth2011, Yang2014, Kopparapu2017, Turbet2018, Wolf2020}. Furthermore, recent investigations have explored the possibility of climate bistability or multistability near this inner edge \citep{Turbet2023}, highlighting the necessity of considering not only the current insolation of a planet, but also its temporal evolution when the long-term habitability is assessed. In contrast, fewer studies have addressed the outer edge of the habitable zone, likely because only a limited number of known planets are situated in this regime. Nonetheless, \citet{Ramirez2017, Ramirez2018} investigated the evolution of the outer boundary in the habitable zone as a function of atmospheric composition using 1D climate models.
    \par
    
    For M-dwarf stars with effective temperatures of about 3400 K, as is the case for Gl725\,B, the conservative inner edge of the habitable zone is generally situated at a stellar flux of approximately 0.8S$_\oplus$, regardless of historical insolation trends \citep{Turbet2023}. The outer boundary, assuming a classical N$_2$-CO$_2$-H$_2$O atmospheric composition, is situated around 0.25S$_\oplus$. This limit can vary slightly depending on the atmospheric makeup: A hydrogen-rich atmosphere (e.g., from volcanic outgassing) can push the boundary outward, while methane-rich atmospheres tend to shift it inward \citep{Ramirez2017}.  The outer edge, assuming a canonical N$_2$–CO$_2$–H$_2$O atmospheric composition, lies near 0.25S$_\oplus$. This limit might be modulated by alternative atmospheric constituents: Enhanced H$_2$ abundances (e.g., via volcanic outgassing) can extend the outer edge outward, whereas elevated CH$_4$ levels might induce a modest inward shift \citep{Ramirez2017}. In all plausible scenarios, Gl725\,Bc remains well ensconced within the habitable zone, however.
    \par
    
    The source Gl725\,Bc is therefore the second closest known Earth or super-Earth-type planet located within the habitable zone of its star after Proxima b. Although its true mass is unknown, we can estimate its most probable value by applying the average correction factor for inclination ($\sqrt{4/3} \sim 1.15$), yielding an estimated mass of ~4 M$_\oplus$. According to \citet{Otegi2020}, this mass range strongly suggests a rocky, terrestrial composition.
    
    \par
    
    Atmospheric characterization of exoplanets is commonly performed through transmission spectroscopy, which requires transiting planets. This is an unlikely configuration for Gl\,725\,Bb and Gl\,725\,Bc. Alternative techniques have been suggested to study the atmospheres of nontransiting nearby exoplanets, however. One promising approach combines high spectral and spatial resolution and is known as high-dispersion coronagraphy (HDC; \citealt{Lovis2017, Snellen2019}).
    
    \par 
    Using the formula from \citet{Lovis2017} and assuming geometric albedos in the range \(A_g = 0.1\)–0.6 and planetary radii of \(1.1\)–\(1.4\,R_\oplus\) for Gl\,725\,Bb and \(1.4\)–\(2.1\,R_\oplus\) for Gl\,725\,Bc, we estimated that the planet-to-star flux ratio lies between \(6 \times 10^{-8}\) and \(3.3 \times 10^{-6}\) for Gl\,725\,Bb and between \(5 \times 10^{-8}\) and \(3.8 \times 10^{-6}\) for Gl\,725\,Bc. Their maximum angular separations, derived from Table~\ref{gl725b_orbital_parameters}, are 9.1 and 39.7\,mas. Based on the diffraction limit of the ELT at 1300\,nm ($2\lambda/D \approx 13.7$\,mas; \citealt{gj1002}), Gl\,725\,Bc would in principle fall within reach of HDC with a next-generation high-resolution spectrograph such as ANDES \citep{Marconi2022}, which will operate in the 400–1800\,nm range at R~$\sim$~100,000. Compared to Proxima\,b, the Earth-like planet in the habitable zone for which reflected-light characterization is currently considered the most accessible, the maximum angular separation of Gl725\,Bc is slightly more favorable. The star–planet contrast might be up to twice as high, however, depending on the actual albedos of the two planets. This makes Gl725\,Bc a particularly valuable target for advancing our understanding of the diversity of planets within the habitable zone.
    
    \par 
        
    Unfortunately, Gl\,725\,B lies in the northern hemisphere and is not accessible to southern facilities such as the ELT. The study of Gl\,725\,Bc would therefore require a comparable HDC instrument located in the north. The peculiar configuration of the BERV excursion for this star means that atmospheric characterization would be optimized using an instrument with a resolving power greater than 70,000 to enable a proper telluric correction. Alternatively, this might be conducted from space.
    
    \par 
    
    Space-based missions, on the other hand, offer an attractive alternative because they are not subject to telluric absorption. The proposed LIFE mission \citep{life}, a mid-infrared nulling interferometer, aims to detect and characterize the atmospheres of Earth-sized and super-Earth exoplanets through their thermal emission. LIFE would consist of four 2-meter apertures operating in the 4–18.5,$\mu$m range and would be particularly well suited to characterizing a planet such as Gl\,725\,Bc. 
\section*{Data availability}

    The data used in this work were recorded in the context of the SLS and are available to the public at the Canadian Astronomy Data Center. {The analysis of these data is available in a \texttt{wapiti} tutorial at \url{https://github.com/HkmMerwan/wapiti}}.
     
    \begin{acknowledgements}

    Based on observations obtained at the Canada-France-Hawaii Telescope (CFHT) which is operated from the summit of Maunakea by the National Research Council of Canada, the Institut National des Sciences de l'Univers of the Centre National de la Recherche Scientifique of France, and the University of Hawaii. The observations at the Canada-France-Hawaii Telescope were performed with care and respect from the summit of Maunakea which is a significant cultural and historic site. Based on observations obtained with SPIRou, an international project led by Institut de Recherche en Astrophysique et Plan\'etologie, Toulouse, France. 
    \par
    This study has ben partially supported through the grant EUR TESS N°ANR-18-EURE-0018 in the framework of the Programme des Investissements d'AveNIR.
    
    \par 
    
    This research made use of the following software tools: \texttt{NumPy}; a fundamental package for scientific computing with Python. \texttt{pandas}; a library providing easy-to-use data structures and data analysis tools.  \texttt{SciPy}; a library for scientific computing and technical computing. \texttt{matplotlib}; a plotting library for the Python programming language. \texttt{tqdm}; a library for creating progress bars in the command line. \texttt{seaborn}; a data visualization library based on matplotlib. \texttt{wpca}; a python package for weighted principal component analysis. \texttt{samsam}: an adaptive MCMC sampler.
    \end{acknowledgements}

    \bibliographystyle{aa}\bibliography{ref}
    
    \begin{appendix}
    
    \section{The \texttt{wapiti} method step-by-step}\label{section:wapiti_step}
    
\begin{figure}
        \centering
        \includegraphics[width=\linewidth]{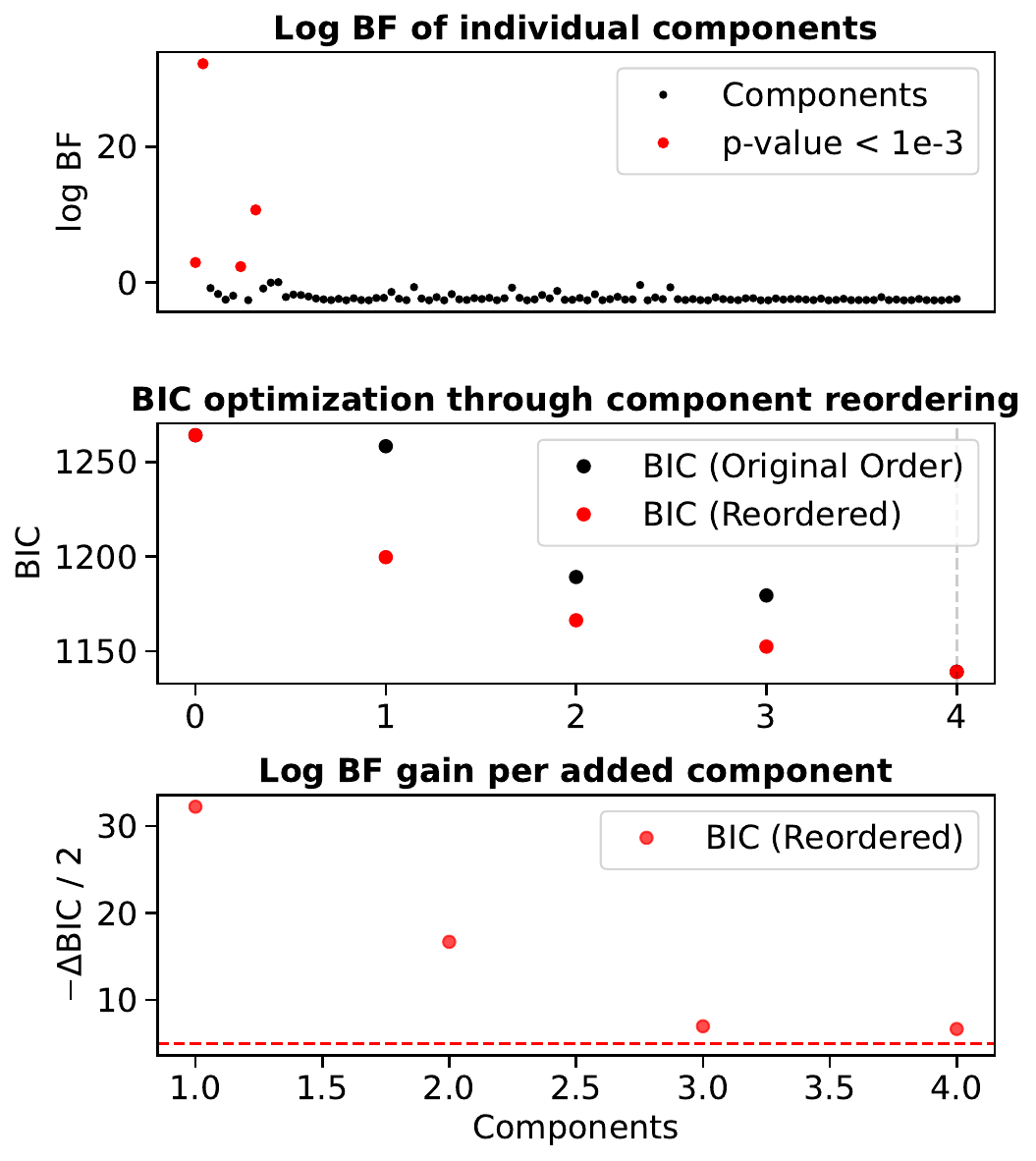}
     \caption{\textit{Top}: log BF of each individual wPCA component fitted separately to the RV data. Components that pass the permutation test with a p-value below $10^{-3}$ are highlighted in red. \textit{Middle}: BIC values before and after reordering the significant components. \textit{Bottom}: Gain in model evidence associated with each additional component and the red dashed line marks the threshold of 5.}
    \label{fig:wapiti_step_by_step}
    \end{figure}

    In this appendix we present the \texttt{wapiti} method in detail, outlining each step in Fig. \ref{fig:wapiti_step_by_step} as applied to the RV data of Gl\,725\,B and a tutorial can be found online\footnote{https://github.com/HkmMerwan/wapiti}. In summary, the \texttt{wapiti} method applies a wPCA to the LBL per-line RV time series, with the goal of finding the principal vectors that best serve as proxies for the systematic errors in the RV data. 
    
    \par 
    
    Before applying wPCA to the data, per-line RV time series with more than 50\% of missing data are excluded. Subsequently, the average RV time series is subtracted from each per-line RV time series to ensure that Keplerian or any achromatic signals, such as trends, remain unaffected by wPCA and are not misinterpreted as systematic errors. Finally, the time series are standardized by using their inverse-variance weighted mean and variance \citep{wapiti}. 
    
    \par 
    
    The \texttt{wapiti} method selects the relevant components by evaluating their contribution to the RV fit. Given a baseline model (e.g., a simple constant offset), we compute the BIC for this model and then the BIC after adding each principal component individually. The log BF comparing these two models allows us to quantify the improvement in fit. The robustness of the log BF value is assessed through a permutation test, in which each column of the RV component data is permuted, effectively shuffling the data to test the stability of the log BF under random configurations. 
    
    \par 
    
    Specifically, for each permutation, a new wPCA model is fitted to the permuted data, and the log BFs are recalculated for each component. This process is repeated 1,000 times, and the resulting distribution of log BFs is examined to determine the significance of the original fit for each component using a p-value test. A component is selected if its log BF value has a p-value below $10^{-3}$. In the case of Gl\,725\,B only the 1st, 2nd, 7th and 9th components are selected (see Fig. \ref{fig:wapiti_step_by_step}).
    
    \par 
    
    Subsequently, we rearrange the selected components in descending order of their BIC values. During this process, we identify the component that, when included, minimizes the BIC, and designate it as the new primary component. We then iteratively determine the subsequent components: at each step, we add the next component that, when combined with the already selected components, further minimizes the BIC. This procedure produces a reordered sequence of principal vectors, denoted $\hat{V_i}$. Using the mathematical notation of \citet{yarara2}, this approach corresponds to calculating the BIC when fitting the following model to the data:
    
    \begin{equation}
       RV(t) = \gamma_{SPIRou} + trend \cdot t + \sum_{i=1}^{N}a_i V_i.
    \end{equation}Here, we include an offset term $\gamma_{SPIRou}$, a trend term (due to the binary companion Gl\,725\,A), the epochs of observation $t$ and the coefficients $a_i$ to fit the principal vectors $V_i$ in the model. The parameter $N$ represents the number of components used in the reconstruction, and can be set to 0, corresponding to the data without applying the \texttt{wapiti} correction. We fit these models using weighted least-squares regression with the \texttt{statsmodels} module \citep{seabold2010statsmodels}, which also provided their BIC values.
    \par 
    
    This reordering method is similar to the $\chi^2$ reordering technique employed in \citet{cameronpca}, the key difference being our use of the BIC as the criterion to penalize the inclusion of too many components. For Gl\,725\,B, the components in BIC-reordered order are the 2nd, 9th, 1st and 7th. To determine how many components to include, we add the BIC-reordered components one by one and monitor the gain in model quality by computing the change in BIC between consecutive models. We stop adding components once the log BF gain per added component drops below a threshold of 5 (see Fig. \ref{fig:wapiti_step_by_step}).
    
    \newpage
    \section{Physical origin of the first principal vector in the case of telluric contamination}\label{section:principal_vector}
    
    To understand the origin of the first principal vector $V_1$, we calculated what we refer to as an indicator of telluric contamination, denoted as $d1_\oplus$. This indicator is determined by computing the Pearson correlation for all lines between the telluric template and the first derivative of the star template. As a reminder, in the LBL algorithm, the RV is determined by subtracting the star template from the observed spectrum and then projecting the residuals onto the first derivative of the star template \citep{etiennelbl}. Therefore, the use of Pearson correlation is a natural choice since it allows us to investigate the linear correlation of this derivative with the telluric lines. By averaging these individual time series, we derive the telluric indicator $d1_\oplus$.
    
    \par 
    
    However, calculating this correlation for all lines would be unnecessary since not all lines contribute to the systematic effects with the same importance. To identify the lines most responsible for the systematics, we apply the method introduced in \citet{wapiti}, which consists in computing the z-scores $z_i$ for each component $i$ among all lines. Using the mathematical framework from \citet{yarara2} this means performing a linear fit of each spectral line $j$ to the principal vectors $V_i$ and obtaining the corresponding coefficients $a_{i, j}$ such as 
    
    \begin{equation}
    RV\left(t, \lambda_j\right) = \gamma_j + \sum_i a_{i, j}V_i\left(t\right),
    \end{equation}where $\gamma_j$ is a constant offset. Using these coefficients, we can then compute the z-scores $z_{i, j}$ of each line $j$ for each component $i$, defined as $\frac{a_{i, j} - \mu_i}{\sigma_i}$, where $\mu_i$ and $\sigma_i$ represent the weighted mean and standard deviation of the coefficients across the spectral lines. 
    
    \par 
    
\begin{figure*}
        \centering
        \includegraphics[width=\linewidth]{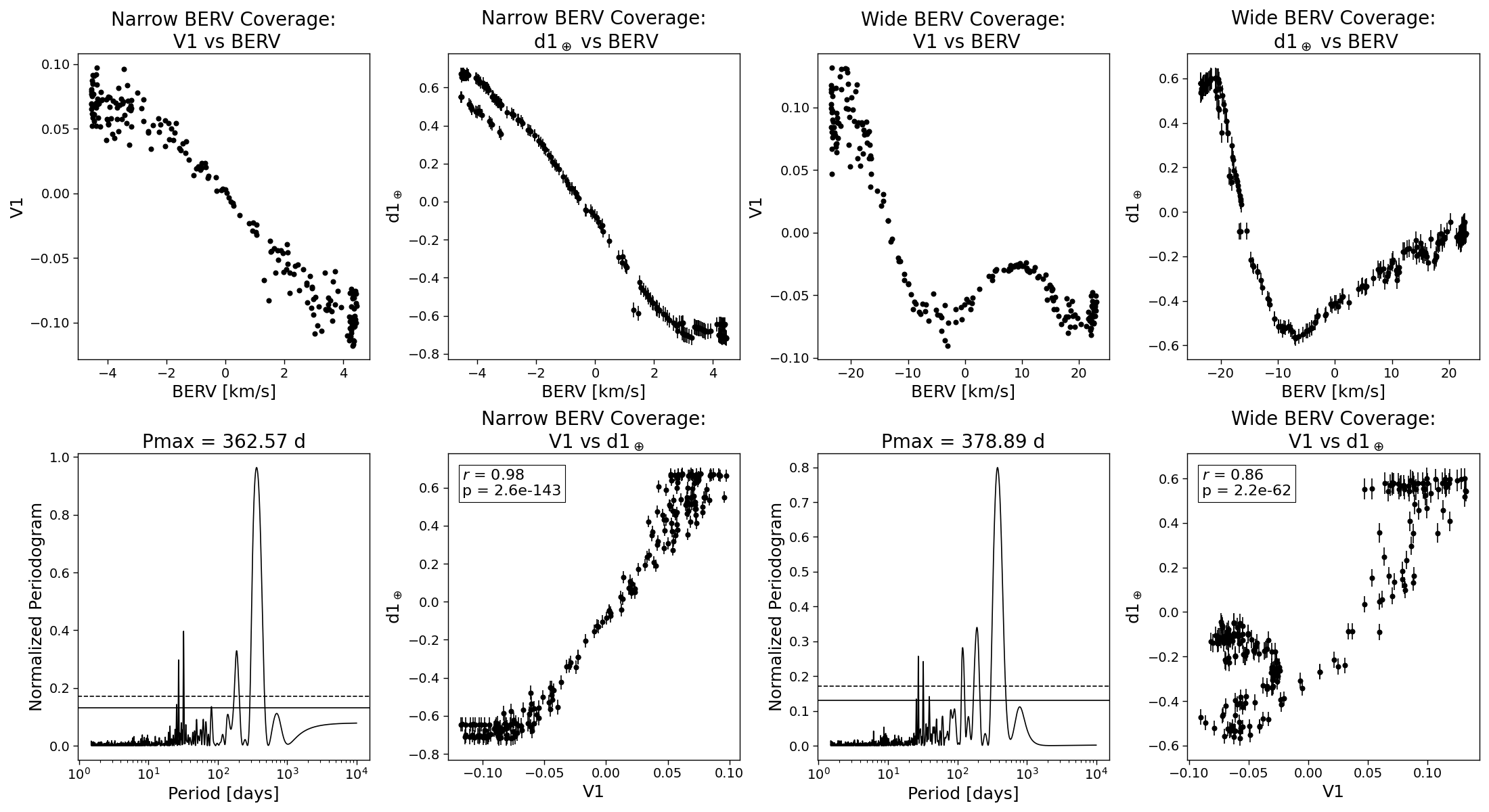}
     \caption{Relationship between the first principal component $V_1$ and the telluric contamination indicator $d1_\oplus$ in the simulated data. \textit{Top}: How $V_1$ and $d1_\oplus$ vary with BERV for narrow and wide coverage. \textit{Bottom}: Periodogram of $V_1$ and the correlation between $V_1$ and $d1_\oplus$.}
    \label{fig:V1_interpretation}
    \end{figure*}

    Consequently, in the two cases of a narrow and wide BERV coverage, we computed $d1_\oplus$ by only calculating the Pearson coefficients for the top 1\% of lines based on the absolute $z_1$ values. We chose the top 1\% lines to fasten the computation but using other values such as 5 or 10\% had no impact on the conclusion of our results. Figure \ref{fig:V1_interpretation} shows that $d1_\oplus$ and $V_1$ exhibit a similar pattern and are, in fact, highly correlated with a Pearson correlation coefficient of $r = 0.98$ and $r = 0.86$ for, respectively, the narrow and wide BERV coverage cases. In conclusion, our interpretation of $V_1$ from these simulations is that it serves as a proxy to assess the impact of telluric absorption lines on the first derivative of the template. 
    
    \section{Telluric contamination of stellar activity indicators}\label{section:contamination_stellar_activity}
    
\begin{figure*}
        \centering
        \includegraphics[width=\linewidth]{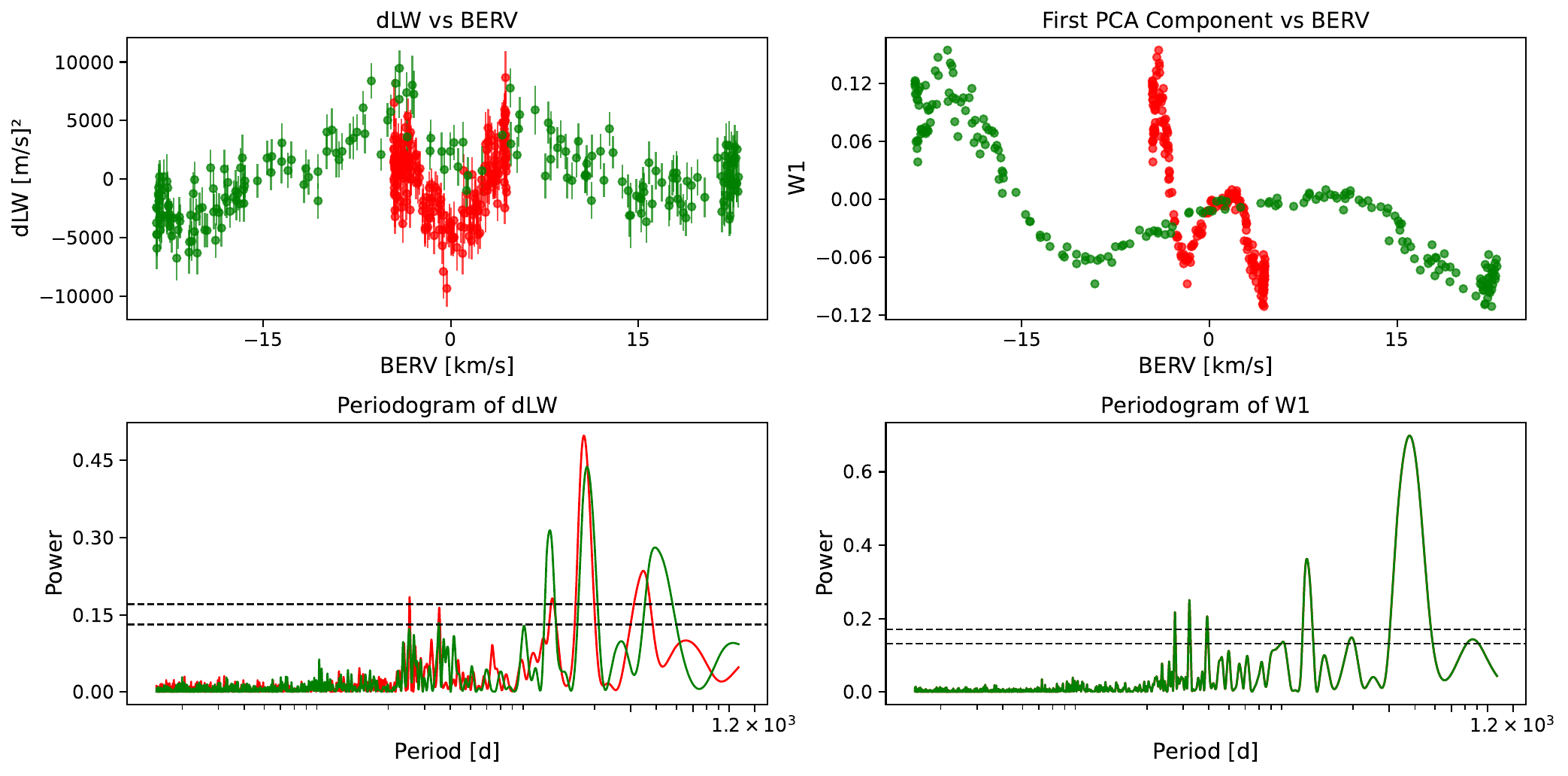}
     \caption[Telluric contamination of the stellar activity indicators]{Telluric contamination of stellar activity indicators of the simulated data in the wide (green) and narrow (red) BERV coverage cases   . \textit{Left}: dLW versus the BERV and the respective periodograms below. \textit{Right}: $W_1$ time series versus the BERV and the respective periodograms below.}
     \label{fig:stellar_contamination}
    \end{figure*}

\begin{figure*}
        \centering
        \includegraphics[width=\linewidth]{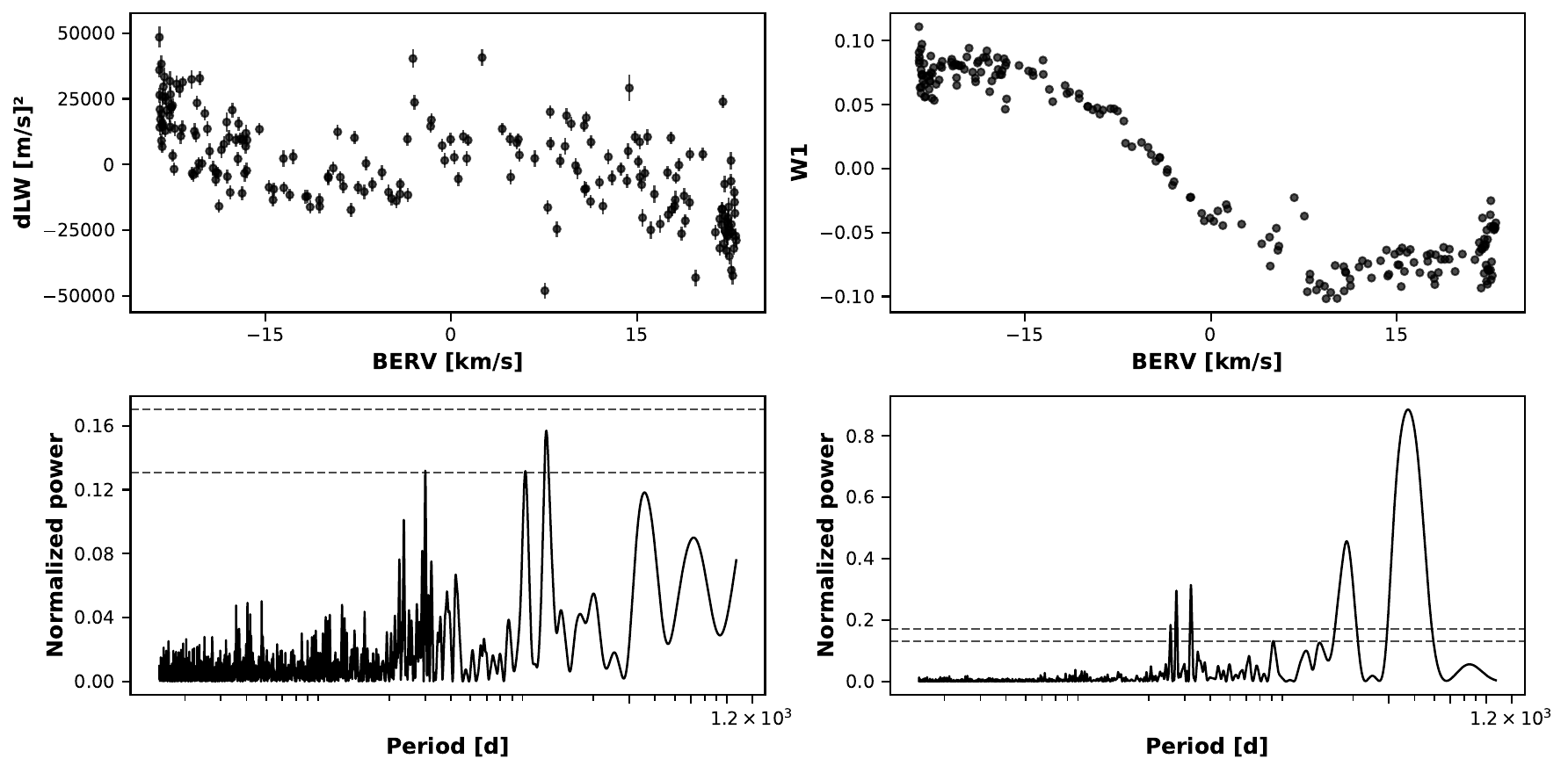}
     \caption[Telluric contamination of the stellar activity indicators]{Telluric contamination of stellar activity indicators of Gl\,725\,B. \textit{Left}: dLW versus the BERV and the respective periodograms (below). \textit{Right}: $W_1$ time series versus the BERV and the respective periodograms.}
     \label{fig:real_stellar_contamination}
    \end{figure*}

    Our simulations also show that stellar activity indicators can be contaminated by tellurics. This is notably the case for the dLW (\citealt{serval}): the dLW quantifies alterations in the average width of line profiles in relation to the template \citep{etiennelbl}. In our simulations, we observed that at the 3\% level of telluric contamination, this indicator is affected in both a wide and narrow BERV coverage (see Fig. \ref{fig:stellar_contamination}). This is also evident in the real observations of Gl\,725\,B, which show similar contamination (Fig. \ref{fig:real_stellar_contamination}). This could pose challenges when attempting to identify and characterize stellar activity within the data, as it compromises the reliability of dLW as a stellar activity indicator.
    
    \par 
    
    Moreover, it is possible to extend the application of wPCA to the per-line dLWs, resulting in the computation of the first principal vector, denoted as $W_1$. This principal vector has shown promise as a stellar activity indicator, as demonstrated when studying the young active M dwarf AU Mic \citep{aumic}. However, $W_1$ also appears to be susceptible to telluric contamination, as seen in both our simulations (Fig. \ref{fig:stellar_contamination}) and real observations (Fig. \ref{fig:real_stellar_contamination}). Thus, the $W_1$ indicator is also vulnerable to tellurics and caution should be exercised when using it for activity characterization.
    
    \par 
    
    If stellar activity creates variability in the per-lines dLWs, it would be more advisable to check for activity signals in all the relevant principal vectors $W_i$ instead of just $W_1$. In the case of AU Mic, its success could be attributed to the high activity level of this young M dwarf, surpassing the influence of tellurics and contributing to the indicator's effectiveness by being responsible for the most variability in the data. Nevertheless, the impact of telluric lines on $W_1$ has been shown to be manageable through a filtering technique that effectively removes telluric contamination, making it a reliable proxy for the small-scale magnetic field \citep{Charpentier2025}.
    
    \section{Supplementary material regarding the RV analysis}

  \subsection{Corner plots of the posterior distributions}\label{section:mcmc_gl725b}

\begin{figure*}
    \centering
    \includegraphics[width=\linewidth]{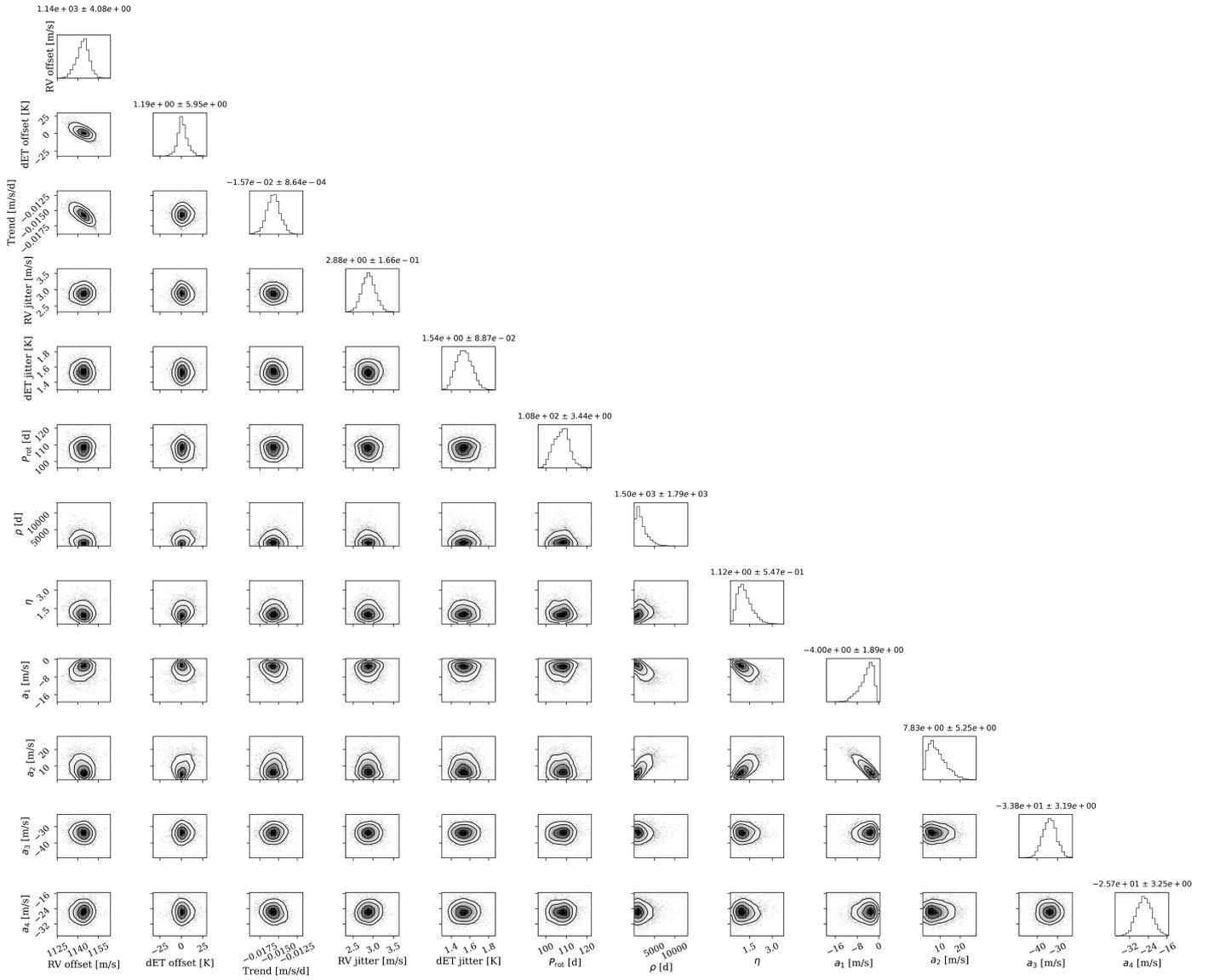}
\caption{Corner plot of the posterior distributions for instrumental, stellar activity, and systematics parameters, including SPIRou offsets, jitter terms, wapiti coefficients, and hyperparameters such as rotation period (\(P_{\mathrm{rot}}\)), timescale (\(\rho\)), and decay term (\(\eta\)).}
\label{fig:mcmc_parameters}
\end{figure*}

\begin{figure*}
    \centering
    \includegraphics[width=\linewidth]{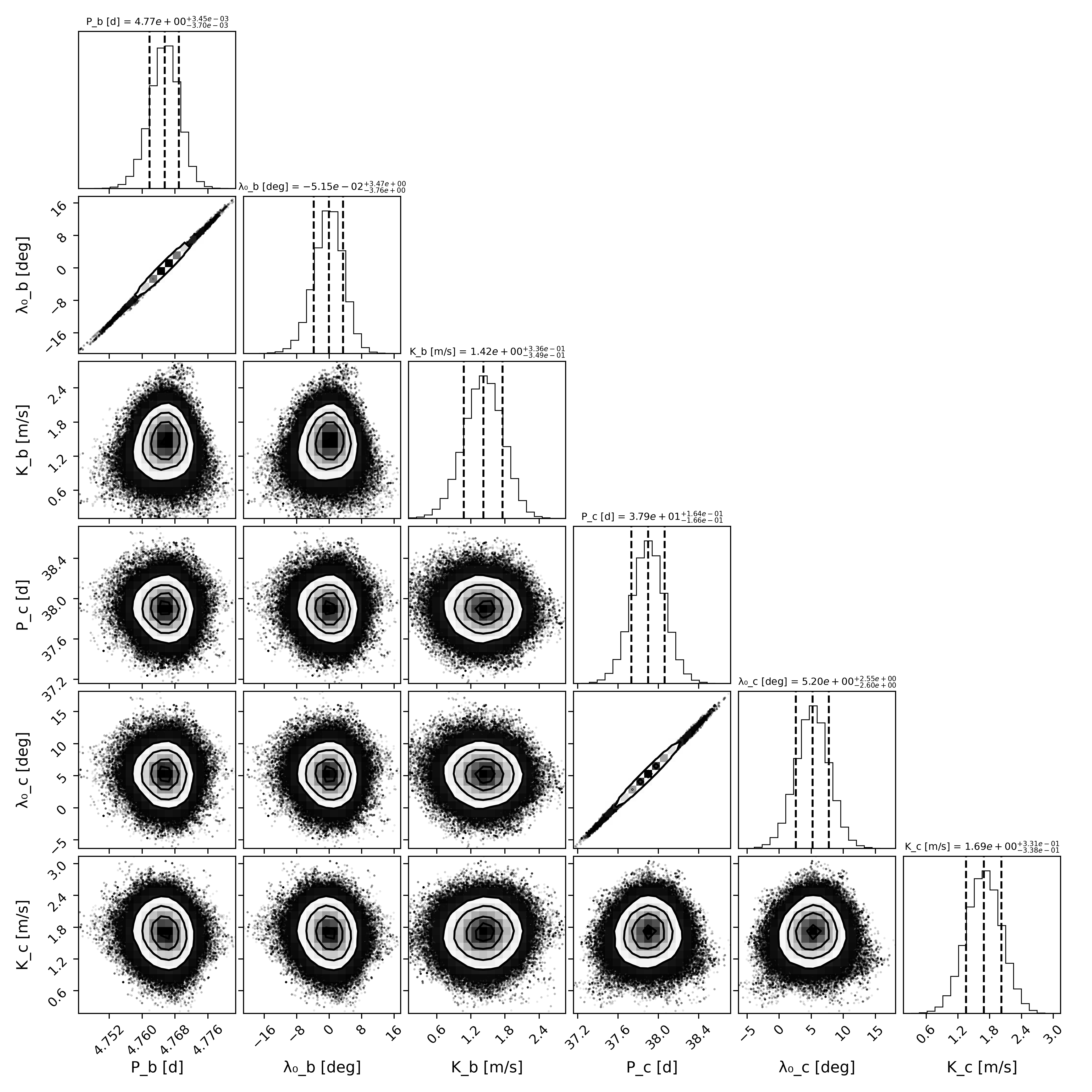}
\caption{Corner plot of the posterior distributions for the orbital parameters of planets b and c, including orbital period, phase (\(\lambda_0\)) and semi-amplitude (\(K\)).}
\label{fig:mcmc_parameters_planet_b}
\end{figure*}

See Figs. \ref{fig:mcmc_parameters} and \ref{fig:mcmc_parameters_planet_b}.

\subsection{Statistical significance of the planetary signals}\label{section:fip}

{To assess the reliability of the two signals detected at $4.77\,$d and $37.9\,$d, we computed a FIP periodogram \citep{fip} from the samples produced by the \texttt{UltraNest} sampler \citep{buchner2021ultranestrobustgeneral} used to estimate the Bayesian evidences of the $n_{\max}=0,1,2,3$–planet models.}

\medskip

\noindent\textbf{1. Sampler and priors.}  
{We employ \texttt{ReactiveNestedSampler} from UltraNest, with a mixed-random \texttt{SliceSampler} step, to explore the joint posterior of the parameters of the model $M_0$ from Eq. \ref{multiGP}:}
{\[
\{\gamma,\ \mathrm{trend},\ (a_i)_{i=1\ldots n_{\rm sys}},\ (A_j,B_j,P_j)_{j=1\ldots n_{\max}}\}.
\]}
{The prior distributions are:}
\begin{itemize}
  \item {$\gamma\sim\mathcal{U}(500,1500)\,\mathrm{m\,s^{-1}}$}
  \item {$\mathrm{trend}\sim\mathcal{U}(-0.1,0.1)\,\mathrm{m\,s^{-1}\,d^{-1}}$,}
  \item {$a_i\sim\mathcal{U}(-100,100)\,\mathrm{m\,s^{-1}}$,}
  \item {$A_j,B_j\sim\mathcal{N}(0,5)\,\mathrm{m\,s^{-1}}$}
  \item {$P_j\sim\mathrm{LogUniform}(1.5,50)\,\mathrm{d}$.}
\end{itemize}
{For each $n_{\max}\in\{0,1,2,3\}$, we perform three independent runs with 400 live points each, and then pool the weighted samples.}

\medskip

\begin{figure}
    \centering
    \includegraphics[width=\linewidth]{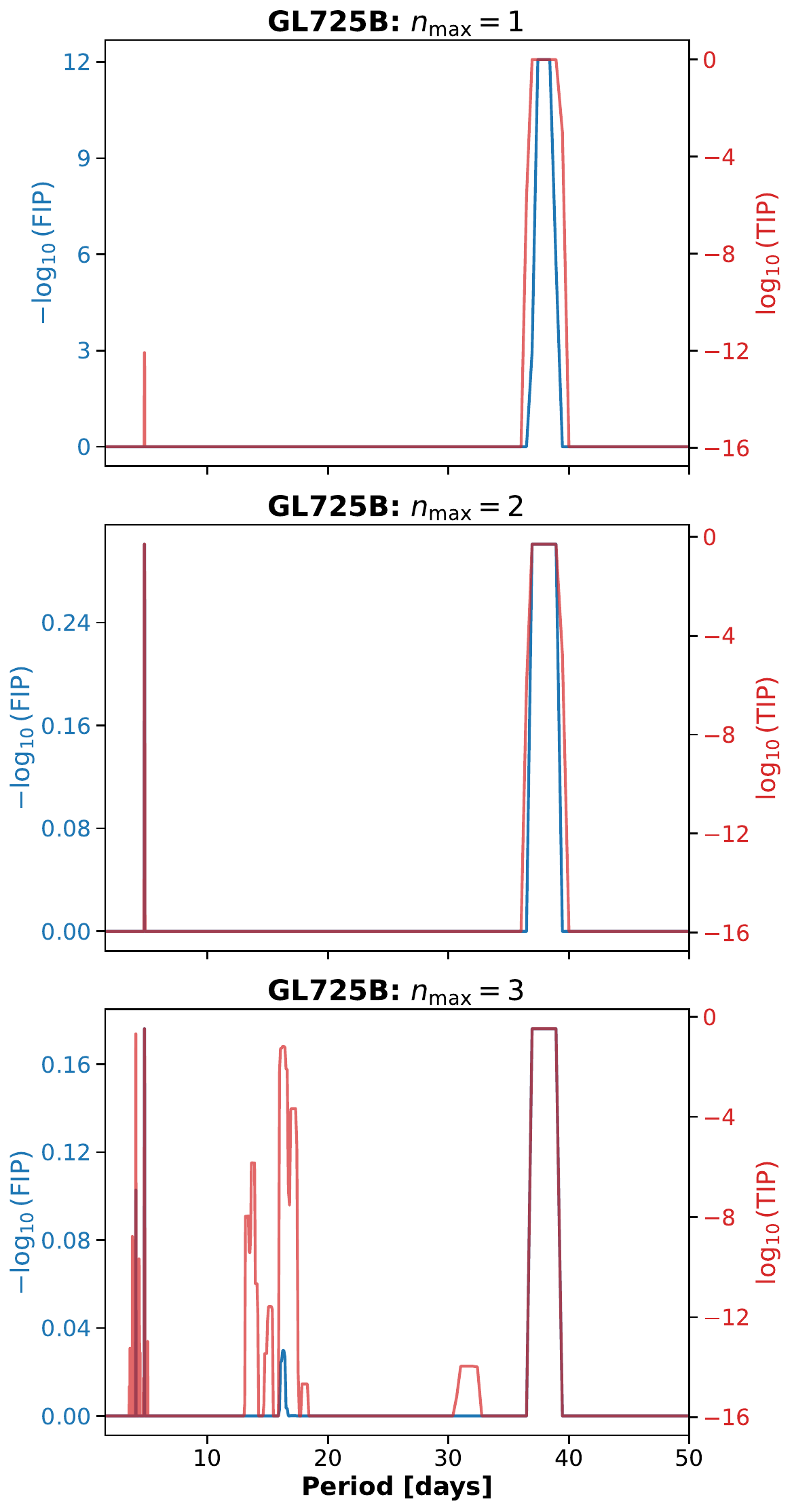}
\caption{FIP periodograms of Gl\,725\,B for models with up to $n_{\max}=1,2,3$ planets (top to bottom).}
\label{fig:GL725B_FIP}
\end{figure}

\noindent\textbf{2. FIP computation and results (Fig.~\ref{fig:GL725B_FIP})}  

{Details on the FIP computation can be found in a tutorial online\footnote{\url{https://github.com/nathanchara/FIP/blob/main/fip_calculation.ipynb}}, we plotted $-\log_{10}\mathrm{FIP}$ (blue) and $\log_{10}\mathrm{TIP}$ (red) as a function of period for $n_{\max}=1,2,3$.}

\begin{itemize}
  \item {The posterior probability to have $n_{\max}=0$ planets is of $5\times10^{-79}$}
  \item T{he posterior probability to have $n_{\max}=1$ planets is of $5\times10^{-45}$, only the $37.9\,$d signal appears and hints of the $4.77\,$d signal appears in the TIP.}
  \item {The posterior probability to have $n_{\max}=2$ planets is of $1\times10^{-20}$, two clear peaks emerge at $4.77\,$d and $37.9\,$d.}
  \item {The posterior probability to have $n_{\max}=3$ planets is of $1$.}
\end{itemize}

\subsection{Sensitivity of the detected signals to the choice of NZP}
\label{app:nzp_sensitivity}

{To assess the robustness of the periodicities against instrumental long-term drifts, we repeated the signal search over a grid of NZP time series. We considered 16 NZP variants and four SNR thresholds applied to the time series (0, 100, 120, 150), yielding $16\times4=64$ \emph{NZP$\times$SNR} realizations of the RV dataset. For each realization, we ran the iterative Bayesian periodogram (Sect.~\ref{sec:725b}), and we recorded the best period and its $\log BF$ at iteration~1. Whenever iteration~1 crossed the detection threshold ($\log BF>5$), we proceeded to iteration~2 and repeated the same procedure on the residuals.}

{For both iterations, we grouped the returned periods into clusters using a 2\% relative tolerance in period and we computed summary statistics per cluster (consensus period, multiplicity across realizations, and evidence metrics). Below we only quote the clusters relevant to the planetary interpretation.}

\paragraph{{Iteration~1 (before adding any Keplerian)}}
Across all \emph{NZP$\times$SNR} realizations, the most frequent first-iteration solution is a cluster centred on
$P = \mathbf{37.894}\,\mathrm{d}$ with a narrow spread (period range $[37.585, 38.051]$\,d) and strong evidence
($\log BF_{\rm mean}=\mathbf{8.05}$, $\log BF_{\rm median}=8.30$, $\log BF_{\rm max}=12.49$; $n=25$ realizations).
Other first-iteration clusters appear less coherent and/or with lower evidence (e.g. a $13.80$\,d cluster with
$\log BF_{\rm mean}\approx0.70$, $n=19$; a $92.40$\,d cluster with $\log BF_{\rm mean}\approx5.19$, $n=5$).
A handful of realizations return a $\sim4.77$\,d peak already at iteration~1 (period range $[4.763,4.771]$\,d,
$\log BF_{\rm max}=8.41$), but with low average support ($\log BF_{\rm mean}\approx-0.21$, $n=5$), indicating that
this short-period solution is not consistently preferred \emph{before} accounting for the $\sim$38\,d signal.

\paragraph{Iteration~2 (conditioned on the iteration~1 detection).}
Among the realizations where iteration~1 was significant (39 in total), the iteration~2 search most frequently returns a cluster at
$P = \mathbf{4.770}\,\mathrm{d}$ (period range $[4.763,4.771]$\,d) with moderate evidence on average
($\log BF_{\rm mean}=\mathbf{1.49}$, $\log BF_{\rm median}=1.29$, $\log BF_{\rm max}=6.76$; $n=23$).
A weaker $\sim$38\,d family is sometimes re-selected at iteration~2 ($\log BF_{\rm mean}\approx2.00$, $n=4$),
as are seasonal/rotational harmonics near $90$--$135$\,d ($\log BF_{\rm mean}\approx4.33$ at $\sim$91\,d; $n=4$).

\begin{figure}
    \centering
    \includegraphics[width=\linewidth]{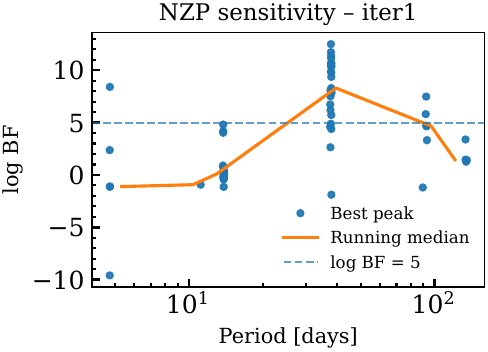}
    \includegraphics[width=\linewidth]{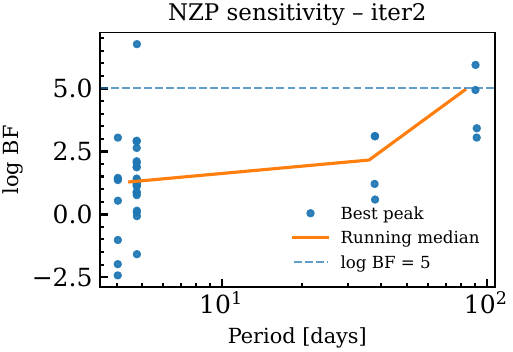}
\caption{\(\log BF\) of the strongest periodic signal as a function of period,
for all tested combinations of NZP realizations and SNR thresholds.
\emph{Top:} Results from the first iteration (before adding any Keplerian component).
\emph{Bottom:} Second iteration, i.e. search in the residuals after the first detected signal is included.
A horizontal dashed line marks the detection threshold at \(\log BF = 5\).}
\label{fig:nzp_sensitivity}
\end{figure}
    
\paragraph{Summary.}
{The $P\!\sim\!38$\,d signal is (i) the most common outcome at iteration~1 across \emph{NZP$\times$SNR} choices, and
(ii) consistently associated with strong $\log BF$ values, with a tight inter-realization period dispersion.
This points to a high stability of the $\sim$38\,d solution with respect to the NZP modelling and supports its planetary nature.}

{Conversely, the $P\!\sim\!4.77$\,d signal becomes the {most frequent} solution {after} the $\sim$38\,d component is included,
but its average evidence at iteration~2 remains modest (though it reaches $\log BF\!\simeq\!6.8$ in the best case). The moderate mean $\log BF$ and the
residual sensitivity to the NZP choice argue for a conservative classification of the $\sim$4.77\,d signal as a candidate.}

    \end{appendix}
\end{document}